\documentclass[sigconf]{acmart}

\usepackage{multirow,balance}
\usepackage{caption}
\usepackage{subcaption}
\usepackage{bm}

\copyrightyear{2020}
\acmYear{2020}
\setcopyright{acmcopyright}\acmConference[RecSys '20]{Fourteenth ACM Conference on Recommender Systems}{September 22--26, 2020}{Virtual Event, Brazil}
\acmBooktitle{Fourteenth ACM Conference on Recommender Systems (RecSys '20), September 22--26, 2020, Virtual Event, Brazil}
\acmPrice{15.00}
\acmDOI{10.1145/3383313.3412238}
\acmISBN{978-1-4503-7583-2/20/09}

\begin{document}

\title{PURS: Personalized Unexpected Recommender System for Improving User Satisfaction}



\author{Pan Li}
\affiliation{%
  \institution{New York University}
  \streetaddress{44 West 4th Street}
  \city{New York}
  \country{USA}}
\email{pli2@stern.nyu.edu}

\author{Maofei Que}
\affiliation{%
  \institution{Alibaba Youku Cognitive and Intelligent Lab}
  \city{Beijing}
  \country{China}}
\email{maofei.qmf@alibaba-inc.com}

\author{Zhichao Jiang}
\affiliation{%
  \institution{Alibaba Youku Cognitive and Intelligent Lab}
  \city{Beijing}
  \country{China}}
\email{zhichao.jzc@alibaba-inc.com}

\author{Yao Hu}
\affiliation{%
  \institution{Alibaba Youku Cognitive and Intelligent Lab}
  \city{Beijing}
  \country{China}}
\email{yaoohu@alibaba-inc.com}

\author{Alexander Tuzhilin}
\affiliation{%
  \institution{New York University}
  \streetaddress{44 West 4th Street}
  \city{New York}
  \country{USA}}
\email{atuzhili@stern.nyu.edu}

\begin{abstract}
Classical recommender system methods typically face the filter bubble problem when users only receive recommendations of their familiar items, making them bored and dissatisfied. To address the filter bubble problem, unexpected recommendations have been proposed to recommend items significantly deviating from user's prior expectations and thus surprising them by presenting "fresh" and previously unexplored items to the users. In this paper, we describe a novel Personalized Unexpected Recommender System (PURS) model that incorporates unexpectedness into the recommendation process by providing multi-cluster modeling of user interests in the latent space and personalized unexpectedness via the self-attention mechanism and via selection of an appropriate unexpected activation function. Extensive offline experiments on three real-world datasets illustrate that the proposed PURS model significantly outperforms the state-of-the-art baseline approaches in terms of both accuracy and unexpectedness measures. In addition, we conduct an online A/B test at a major video platform Alibaba-Youku, where our model achieves over 3\% increase in the average video view per user metric. The proposed model is in the process of being deployed by the company.
\end{abstract}


\maketitle

\section{Introduction}
Recommender systems have been an important tool for online commerce platforms to assist users in filtering for the best content while shaping their interests at the same time. To achieve this objective, collaborative filtering has been among the most popular techniques and was successfully deployed for decades. However as shown in the recent literature, classical collaborative filtering algorithms often lead to the problem of over-specialization \cite{adamopoulos2014over}, filter bubbles \cite{pariser2011filter,nguyen2014exploring} and user boredom \cite{kapoor2015like,kapoor2015just}. To make things worse, users might get annoyed if they are recommended similar items repeatedly in a short period of time.

To address these issues, researchers have proposed to incorporate unexpectedness metric in recommendation models \cite{murakami2007metrics}, the goal of which is to provide novel, surprising and satisfying recommendations. Unlike diversity, which only measures dispersion among recommended items, unexpectedness measures deviations of recommended items from user expectations and thus captures the concept of user surprise and allows recommender systems to break from the filter bubble. It has been shown in \cite{adamopoulos2014discovering,adamopoulos2015unexpectedness} that unexpectedness leads to significant increase of user satisfaction. Therefore, researchers have introduced various recommendation algorithms to optimize the unexpectedness metric and achieved strong recommendation performance \cite{adamopoulos2015unexpectedness,zhang2012auralist,lu2012serendipitous,li2019latent3}.

However, very few of them have been successfully applied to industrial applications or achieved significant improvements in real business settings. This is the case for the following reasons. First, to improve unexpectedness of recommended items, previous models often have to sacrifice the business-oriented metrics \cite{adomavicius2008overcoming,zhou2010solving}, such as CTR (Click-Through-Rate) and GMV (Gross-Merchandise-Volume). Second, most of the proposed unexpected recommendation algorithms are not scalable and therefore cannot be deployed in large-scale industrial applications. Third, the lack of personalization and session-based design of recommender systems reduces user satisfaction and their online browsing performance metrics. Thus, it is important to overcome these problems and intelligently deploy unexpected recommender systems in real-world applications.

Note that, while working on unexpected recommendations, it is crucial to address the problem of heterogeneous user preferences by focusing on providing personalized recommendations according to these preferences. For example, some people tend to be `'variety-seekers'' \cite{mcalister1982variety} and are more willing to click on novel item recommendations, while others prefer to stay in their own comfort zones and are in favor of familiar item recommendations. In addition, even the same person might have different perceptions of unexpectedness in different contexts, which also motivates us to include session-based information into the design of an unexpected recommender system. For example, it is more reasonable to recommend the next episode of a TV series to the user who has just finished the first episode, instead of recommending new types of videos to that person. On the other hand, if the user has been binge-watching the same TV series in one night, it is better to recommend something different to him or her.

To address these concerns, in this paper we propose the \textit{Personalized Unexpected Recommender System} (PURS) that incorporates unexpectedness into recommendations in an end-to-end scalable manner. Specifically, we model user interests as \textit{clusters} of the previously consumed items in the \textit{latent} space and subsequently calculate unexpectedness as the weighted distance between a new item and the clusters of these interests. Furthermore, we utilize sequence modeling and the self-attention mechanism to capture personalized and session-based user perception of unexpectedness. We also propose a novel unexpected activation function to adjust the network output for better unexpected recommendation performance. To provide unexpected recommendations, we construct a hybrid utility function by combining the aforementioned unexpectedness measure with estimated business performance metrics, especially the click-through-rate. Finally, extensive offline experiments on three real-world datasets illustrate that the proposed model consistently and significantly outperforms the state-of-the-art baseline approaches in both accuracy and novelty measures. We also conduct an online A/B test on a major video platform Alibaba-Youku, where our model achieves significant improvements over the current production model. Based on these positive results, the proposed model is in the process of being moved into production in the company.

This paper makes the following research contributions. It

(1) presents a novel PURS model that efficiently and effectively incorporates unexpectedness into recommendations;

(2) proposes to use the self-attention mechanism to model personalized and session-based user perception of unexpectedness;

(3) proposes to calculate unexpectedness as the distance between a new item and clusters of user interests in the latent space;

(4) presents a novel unexpected activation function to optimize performance of unexpected recommendations;

(5) presents extensive offline experiments and an online A/B test that empirically demonstrate the strengths of the PURS model.

\section{Related Work}
In this section, we discuss prior literature related to our work in three categories: unexpected recommendations, deep-learning based recommendations and personalized \& session-based recommendations.

\subsection{Unexpected Recommendations}
As discussed in the previous section, to overcome the problem of filter bubbles and user boredom, researchers have proposed to optimize beyond-accuracy objectives, including unexpectedness, serendipity, novelty, diversity and coverage \cite{murakami2007metrics,ge2010beyond}. Note that, these metrics are closely related to each other, but still different in terms of definition and formulation. Specifically, serendipity measures the positive emotional response of the user about a previously unknown item and indicates how surprising these recommendations are to the target users \cite{shani2011evaluating,chen2019serendipity}; novelty measures the percentage of new recommendations that the users have not seen before or known about \cite{mcnee2006being}; diversity measures the variety of items in a recommendation list, which is commonly modeled as the aggregate pairwise similarity of recommended items \cite{ziegler2005improving}; coverage measures the degree to which recommendations cover the set of available items \cite{herlocker2004evaluating}.

Among them, unexpectedness has attracted particular research interests, for it is shown to be positively correlated with user experience \cite{adamopoulos2015unexpectedness,chen2019serendipity}. It also overcomes the overspecialization problem \cite{adamopoulos2015unexpectedness,iaquinta2010can}, broadens user preferences \cite{herlocker2004evaluating,zhang2012auralist,zheng2015unexpectedness} and increases user satisfaction \cite{adamopoulos2015unexpectedness,zhang2012auralist,lu2012serendipitous}. Unexpectedness measures those recommendations that are not included in the users' previous purchases and depart from their expectations. Different from the diversity measure, unexpectedness is typically defined as the distance between the recommended item and the set of expected items in the feature space \cite{adamopoulos2015unexpectedness}. However, as pointed out in \cite{li2019latent3,li2020latent}, it is difficult to properly construct the distance function in the feature space, thus the authors propose to calculate the distance of item embeddings in the latent space instead. 

Researchers have thus proposed multiple recommendation models to improve novelty measures, including Serendipitous Personalized Ranking (SPR) \cite{lu2012serendipitous} that extends traditional personalized ranking methods by discounting item popularity in AUC optimization; Auralist \cite{zhang2012auralist} that balances the accuracy and novelty measures simultaneously using topic modeling; Determinantal Point Process (DPP) \cite{gartrell2017low,chen2018fast} that utilizes the greedy MAP inference to diversify the recommendation results; HOM-LIN \cite{adamopoulos2015unexpectedness} that use a hybrid utility function of estimated ratings and unexpectedness to provide recommendations; and Latent Modeling of Unexpectedness \cite{li2019latent3,li2020latent}. All of these models have successfully improved the unexpectedness measure of recommendations.

However, these prior literature aim to provide uniform unexpected recommendations to all users under all circumstances, without taking into account user-level heterogeneity and session-based information, thus might not reach the optimal recommendation performance. In addition, those models have the scalability issue and might not work well in the large-scale industrial settings. In this paper, we propose a novel deep-learning based unexpected recommender system to provide personalized and session-based unexpected recommendations in an efficient manner.

\subsection{Deep-Learning based Recommendations}
Another body of related work is around deep-learning based approaches to extract user preferences and provide recommendations. Compared with classical recommender systems, deep-learning based models are capable of conducting representation learning, sequence modeling and nonlinear transformation to possess high level of flexibility \cite{zhang2019deep}. Some representative work include \cite{shi2017survey,shi2018heterogeneous,dong2017metapath2vec} that construct heterogeneous information network embeddings to model complex heterogeneous relations between users and items; \cite{rumelhart1985learning,he2017neural} that apply deep neural network (DNN) techniques to obtain semantic-aware representations of users and items for efficient learning and recommendation; and \cite{hinton2006reducing,sedhain2015autorec} that utilize autoencoder (AE) techniques to map the features of users and items into latent embeddings and model their relationships accordingly.

In this paper, we adopt deep-learning based approaches to model unexpectedness and subsequently provide unexpected recommendations, thus obtaining all these advantages discussed above.

\subsection{Personalized and Session-based Recommendations}
Note that different users might have different preferences towards the same recommendation, and even the same user might have different opinions towards the same recommendations on different sessions. Therefore, personalized and session-based recommender systems constitute important tools for providing satisfying recommendations. Specifically, these models take user behavior sequences into account to learn user behavior patterns and the shift of user preferences from one transaction to another \cite{wang2019survey}. Recent work on this field include DIN \cite{zhou2018deep}, DSIN \cite{feng2019deep}, DIEN \cite{zhou2019deep}, DeepFM \cite{guo2017deepfm}, Wide \& Deep \cite{cheng2016wide} and PNN \cite{qu2016product} that combine user features, item features and user historic behaviors to provide useful recommendations.

In this paper, we incorporate personalized and session-based information into the model design to provide unexpected recommendations and improve user satisfaction.

\section{Unexpectedness}
\subsection{Modeling of Unexpectedness}
In \cite{adamopoulos2015unexpectedness}, the authors define unexpectedness of a recommended item as the distance between the new item and the closure of previous consumptions in the \textit{feature} space. As discussed in the related work section, this approach might not achieve optimal performance, for the distance function is typically difficult to define in the feature space. Therefore in \cite{li2019latent3}, unexpectedness is constructed in the \textit{latent} space instead of the feature space, which is calculated as the euclidean distance to the entire latent closure of previous consumptions. This latent modeling approach of unexpectedness manages to achieve strong recommendation performance and improve novelty of recommendations without losing accuracy measures.

\begin{figure*}
    \centering
    \begin{subfigure}[t]{0.33\textwidth}
        \centering
        \includegraphics[width=\textwidth]{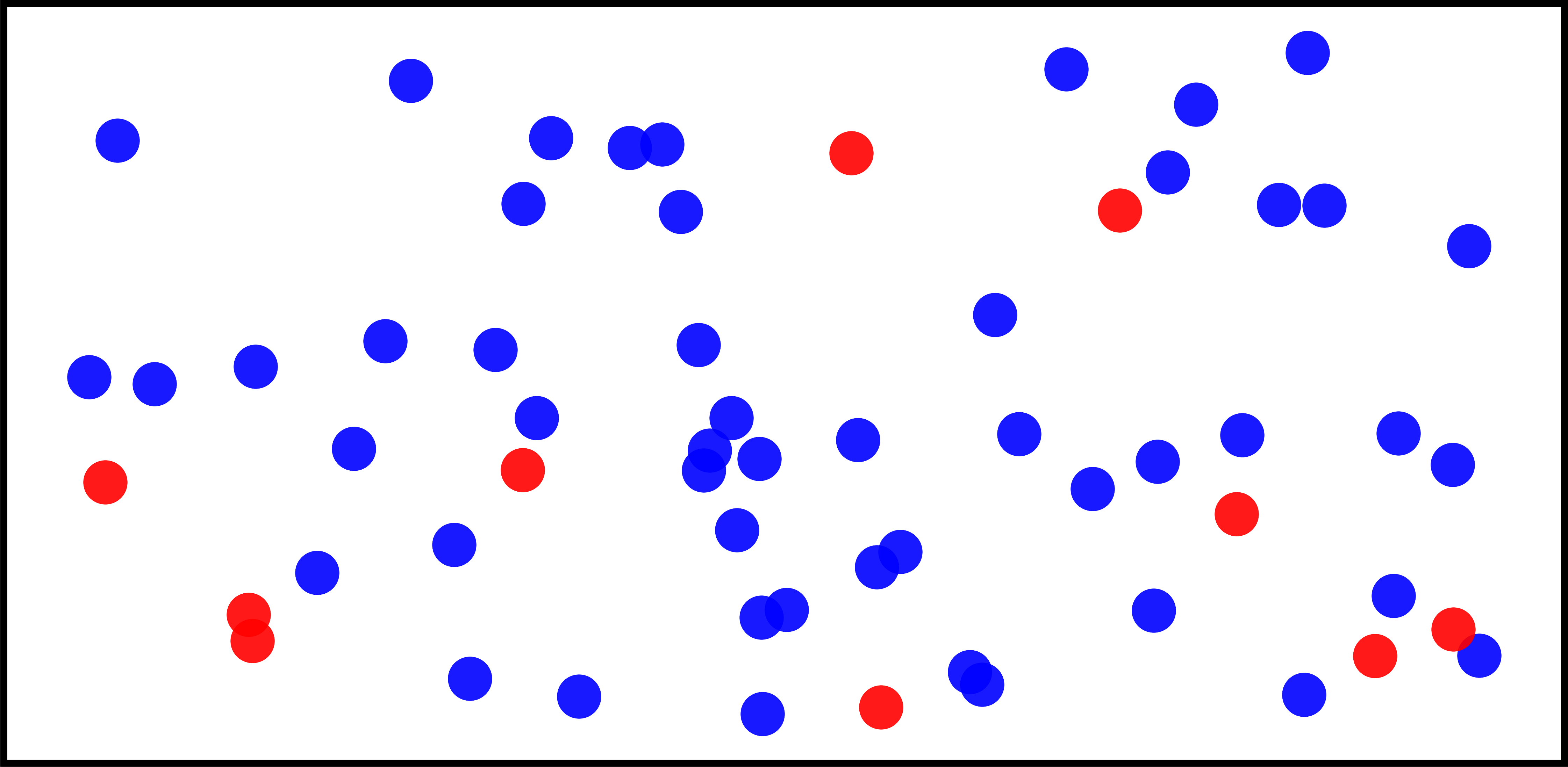}
        \caption{Item Embeddings}
    \end{subfigure}%
    \begin{subfigure}[t]{0.33\textwidth}
        \centering
        \includegraphics[width=\textwidth]{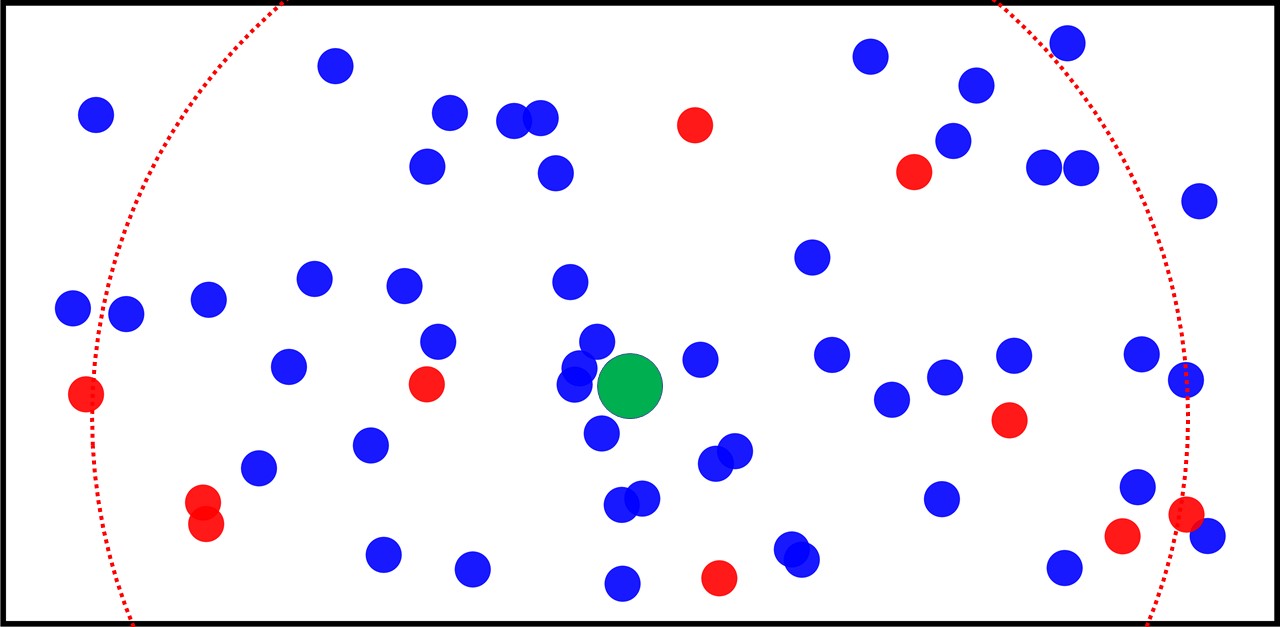}
        \caption{One Single Latent Closure}
    \end{subfigure}%
    \begin{subfigure}[t]{0.33\textwidth}
        \centering
        \includegraphics[width=\textwidth]{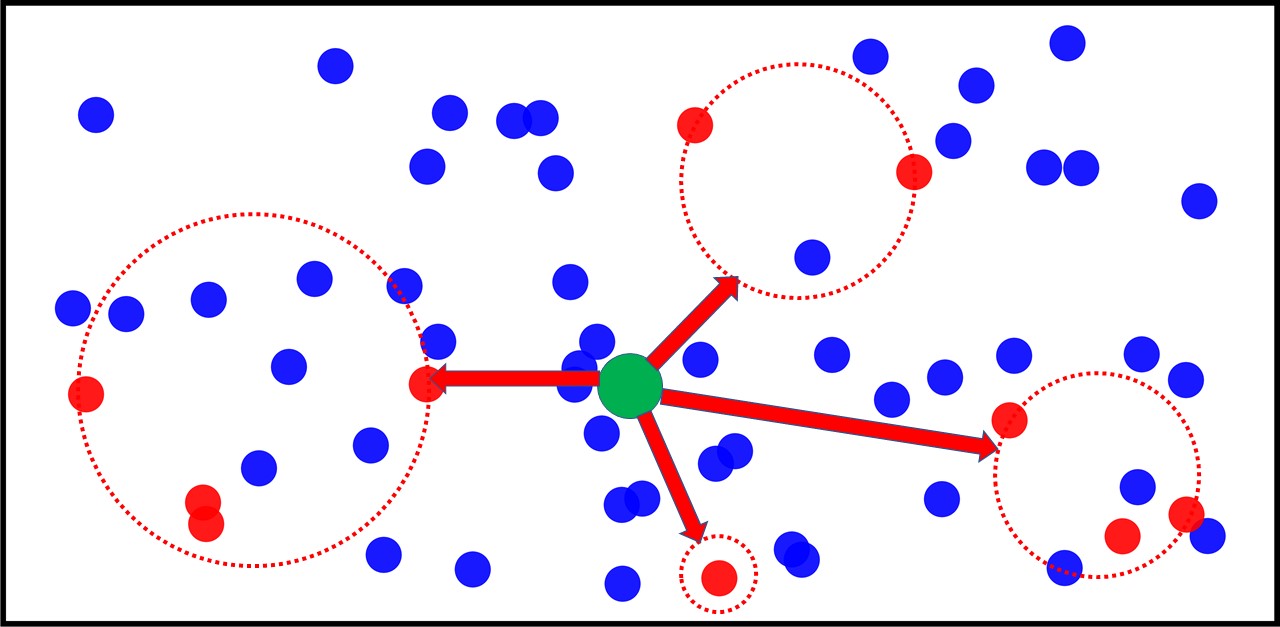}
        \caption{Clustering of Latent Closure}
    \end{subfigure}%
\caption{Comparison of unexpectedness modeling in the latent space. Blue points stand for the available items; Orange points represent the consumed items; Green point refers to the new recommended item. We propose to define unexpectedness as the distance between the new item and clusters of latent closure generated by all previous consumptions.}
\label{definition}
\end{figure*}

Note however, that many users have diverse interests, thus making it unreasonable to use one single closure of consumed items to model user interests. For example, in the movie recommendation applications, a user might have several different types of movie interests, including documentaries, fiction movies, comedies, animation and so on. Therefore, it is suboptimal to use one single closure to model wide interests and might be more appropriate to model user interests in separate clusters instead. In addition, as shown in Figure \ref{definition}, if we take one single closure including all items that the user has consumed as the expected set, it might lead to the construction of an extremely large expected set and might accidentally include some unexpected items into the expected set as well. However, if we cluster the user's historic consumptions based on the diverse interests, it would be easier to identify the behavior patterns within each cluster and model the expectation sets accordingly. It also encourages those unexpected recommendations that could bridge the diverse interests of the same user, as illustrated in Figure \ref{definition}.

Therefore, we propose to conduct clustering on embeddings of previous consumptions and form user interest clusters accordingly. We select the Mean Shift algorithm \cite{cheng1995mean} to identify the clusters of historic behaviors automatically in the latent space for the following reasons. First, it is an unsupervised clustering algorithm, therefore we do not have to explicitly specify the number of interest clusters for each user, as we generally do not have that information as well. Mean Shift algorithm is capable of optimally selecting the best number of clusters without manual specification. Second, it is powerful for the analysis of a complex multi-modal feature space for recommendation applications, and it can also delineate arbitrarily shaped clusters in it \cite{comaniciu2002mean}.

Mean Shift clustering utilizes an iterative process to locate the maxima of a density function given discrete data sampled from that function. To handle the clustering procedure in our model, we denote the kernel function as $K(x_i - x)$ given an initial estimate $x_{i}$ and observation $x$. This kernel function determines the weight of nearby points for re-estimation of the mean, which is typically modeled as a Gaussian distribution
\begin{equation}
K(x_{i}-x)=e^{-c||x_{i}-x||^{2}}
\end{equation}
The weighted mean of the density in the window determined by $K$ is calculated as
\begin{equation}
m(x) = \frac{\sum_{x_i \in N(x)} K(x_i - x) x_i }{\sum_{x_i \in N(x)} K(x_i - x)}
\end{equation}
where $N(x)$ is the neighborhood of $x$. The mean-shift algorithm will reset $x$ as $m(x)$, and repeat the estimation process until $m(x)$ converges.

For each user $u$, we extract the historic behavior sequence as $\{i_{1}, i_{2}, \cdots, i_{n}\}$ and their corresponding embeddings in the latent space as $\{w_{1}, w_{2}, \cdots, w_{n}\}$ through sequence modeling. We subsequently apply Mean Shift algorithm to cluster these embeddings into user interest clusters as $\{C_{1}, C_{2}, \cdots, C_{N}\}$. For each new item recommendation $i_{*}$ for user $u$, unexpectedness is hereby modeled as the weighted average distance between each cluster $C_{k}$ and embedding of the new item $w_{*}$:
\begin{equation}
unexp_{i_{*},u} = \sum_{k=1}^{N} d(w_{*}, C_{k})\times\frac{|C_{k}|}{ \sum_{k=1}^{N} |C_{k}|}
\end{equation}

\subsection{Unexpected Activation Function}
Though it is natural to directly include the unexpectedness obtained from the previous section into the utility function, it is suboptimal to do so, as our goal is to provide unexpected yet relevant and useful recommendations. If we explicitly combine unexpectedness into the hybrid utility function, it will tend to recommend items with very high unexpectedness, which are likely to be either irrelevant or even absurd to the user. In \cite{adamopoulos2015unexpectedness}, the authors propose to use a unimodal function to adjust the unexpectedness input to the utility function. 

However, unimodality alone is not enough for any function to serve as the unexpected activation function, as we also need to balance between unexpectedness and relevance objectives. For example, the unimodal Gaussian function does not obatin the optimal recommendation performance, as shown in Section 6. We consequently propose that the unexpected activation function $f(\cdot)$ should satisfy the following mathematical properties:

\begin{enumerate}
\item \textbf{Continuity} For two items $i_{1}$ and $i_{2}$, if their unexpectedness towards user $u$ are close enough ($|unexp_{u,i_{1}}-unexp_{u,i_{2}}|<\epsilon$), then their unexpectedness output to respective utility should also be close  ($|f(unexp_{u,i_{1}})-f(unexp_{u,i_{2}})|<\delta$), which implies the continuity requirement of $f(\cdot)$. 

\item \textbf{Boundedness} Note that the utility function $utility_{u,i}$ for any user $u$, item $i$ should be bounded, therefore the corresponding unexpectedness output and the unexpected activation function should also be bounded ($|f(\cdot)|<\infty$). In addition, when unexpectedness goes to zero or infinity, it suggests the new item is either too similar to previous consumptions or totally irrelevant, thus its contribution to utility function should be negligible ($lim_{x\to 0} f(x) = 0$, $lim_{x\to \infty} f(x) = 0$).

\item \textbf{Unimodality} For the optimization convenience, it is ideal to have a unimodal unexpectedness output to the utility function instead of multi-peaks which require additional effort for the recommendation model to balance between unexpectedness and relevance, as discussed in \cite{adamopoulos2015unexpectedness}. Therefore, we need to select a unimodal function as the activation function.

\item \textbf{Short-Tailed} To provide unexpected yet relevant recommendations, it is important to note that relevance decreases very fast after unexpectedness increases above certain threshold, which indicates the use of a short-tailed or sub-Gaussian distribution \cite{park2008long}. Specifically, a sub-Gaussian distribution is a probability distribution with strong tail decay, whose tails are dominated by or decay at least as fast as the tails of a Gaussian. Therefore, the activation function should follow the sub-Gaussian distribution, the kurtosis of which should be less than 3 \cite{buldygin1980sub}.
\end{enumerate}

Many functions satisfy the properties above and thus become potential candidates for the unexpected activation function. To simplify the model and accelerate the optimization process, in this paper we choose a popular solution $f(x)=x*e^{-x}$ from the Gamma function \cite{davis1959leonhard} as the unexpected activation function, which satisfies all four required mathematical properties--it is a continuous, bounded and unimodal function with kurtosis value 1.5. As we show in Section 6.3, the unexpected activation function contributes significantly to the superior recommendation performance. 

\begin{figure}
\centering
\includegraphics[width=0.3\textwidth]{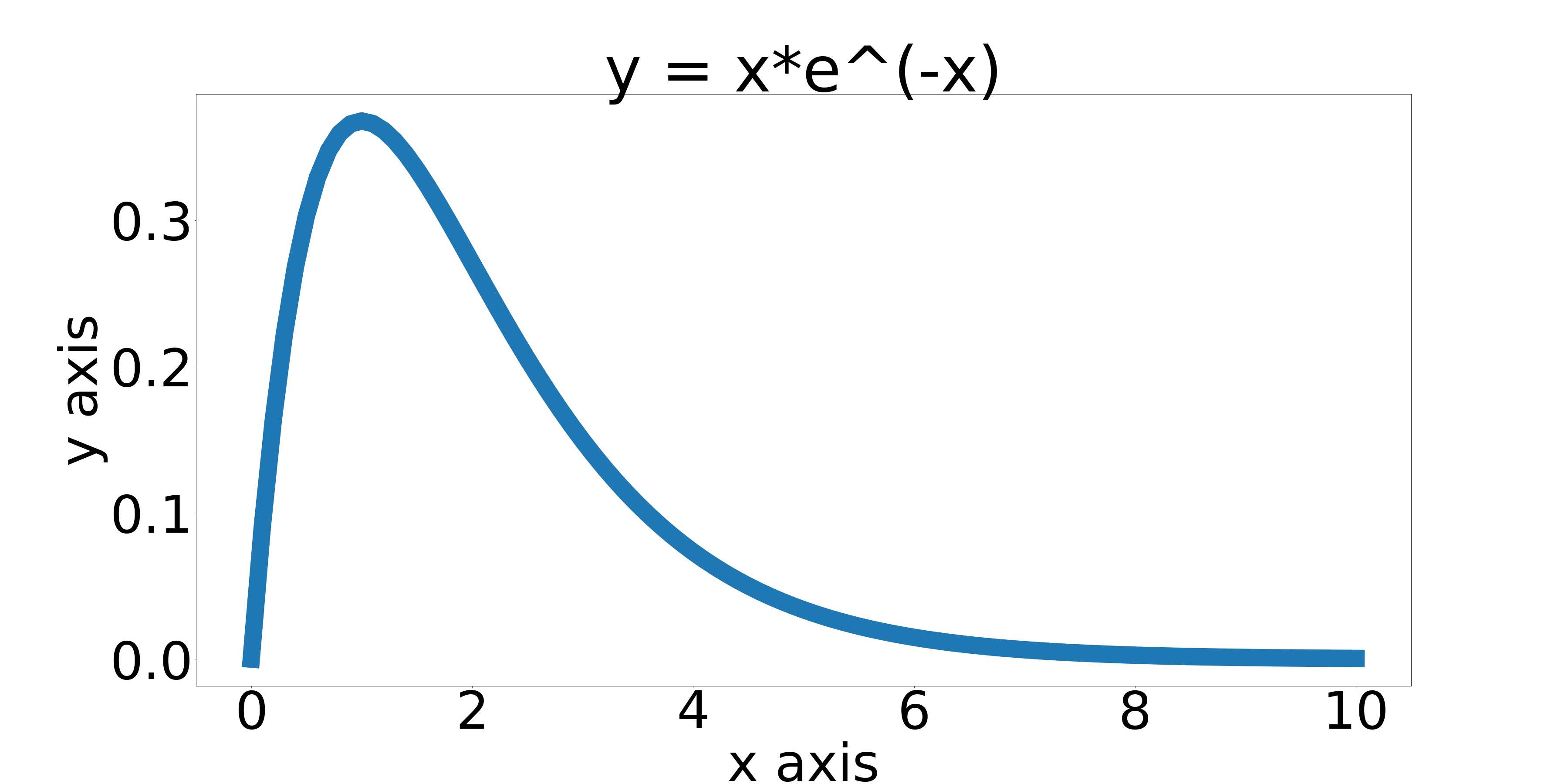}
\caption{Unexpected Activation Function} 
\label{function}
\end{figure}

\section{Personalized Unexpected Recommender System}
\subsection{Overview}
To provide unexpected recommendations, we propose to use the following hybrid utility function for user $u$ and item $i$
\begin{equation}
Utility_{u,i} = r_{u,i} + f(unexp_{u,i})*unexp\_factor_{u,i}
\end{equation}
which consists of the following components:

\bm{$r_{u,i}$} that represents the click-through-rate estimation for user $u$ towards item $i$ based on their features and past behaviors.

\bm{$unexp_{u,i}$} that represents the unexpectedness of item $i$ towards user $u$, as introduced in the previous section.

\bm{$unexp\_factor_{u,i}$} that represents the personalized and session-based perception of unexpectedness for user $u$ towards item $i$. 

\bm{$f(\cdot)$} that stands for the activation function for unexpectedness in order to effectively and efficiently incorporate this piece into the utility function, as introduced in the previous section.

\begin{figure*}
    \centering
    \begin{subfigure}[t]{0.5\textwidth}
        \centering
        \includegraphics[width=\textwidth]{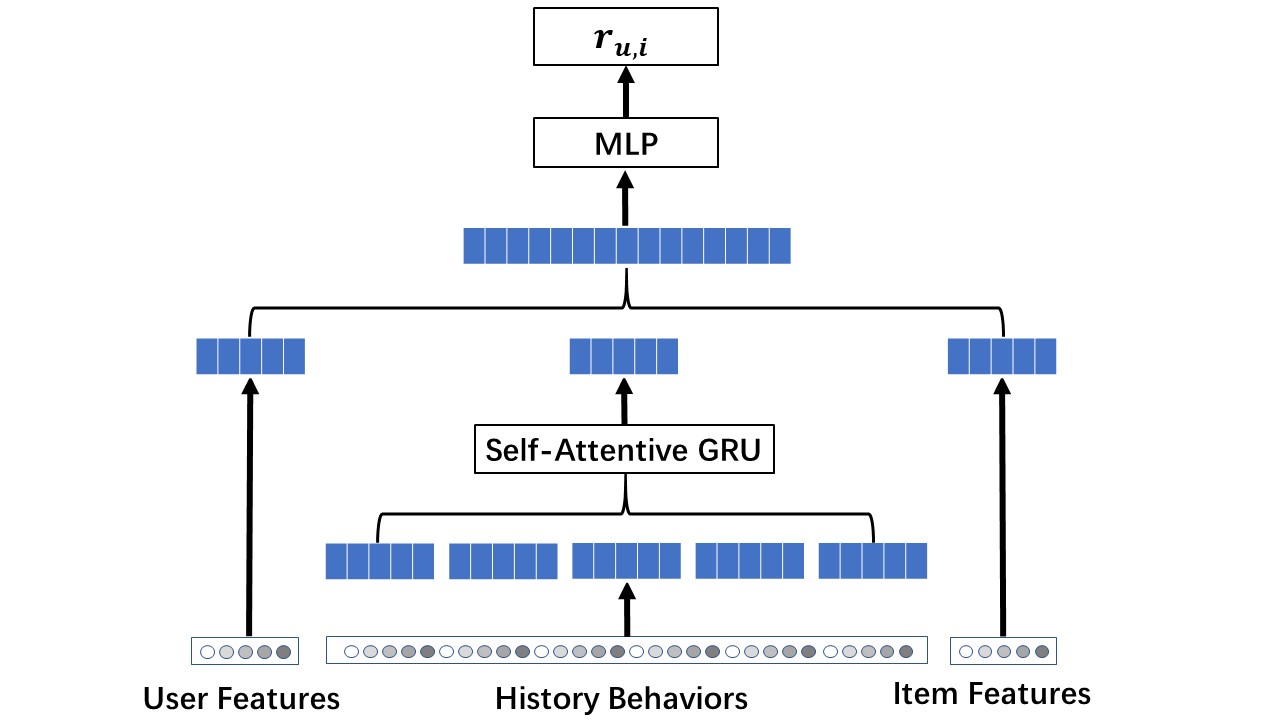}
        \caption{Base}
    \end{subfigure}%
    \begin{subfigure}[t]{0.5\textwidth}
        \centering
        \includegraphics[width=\textwidth]{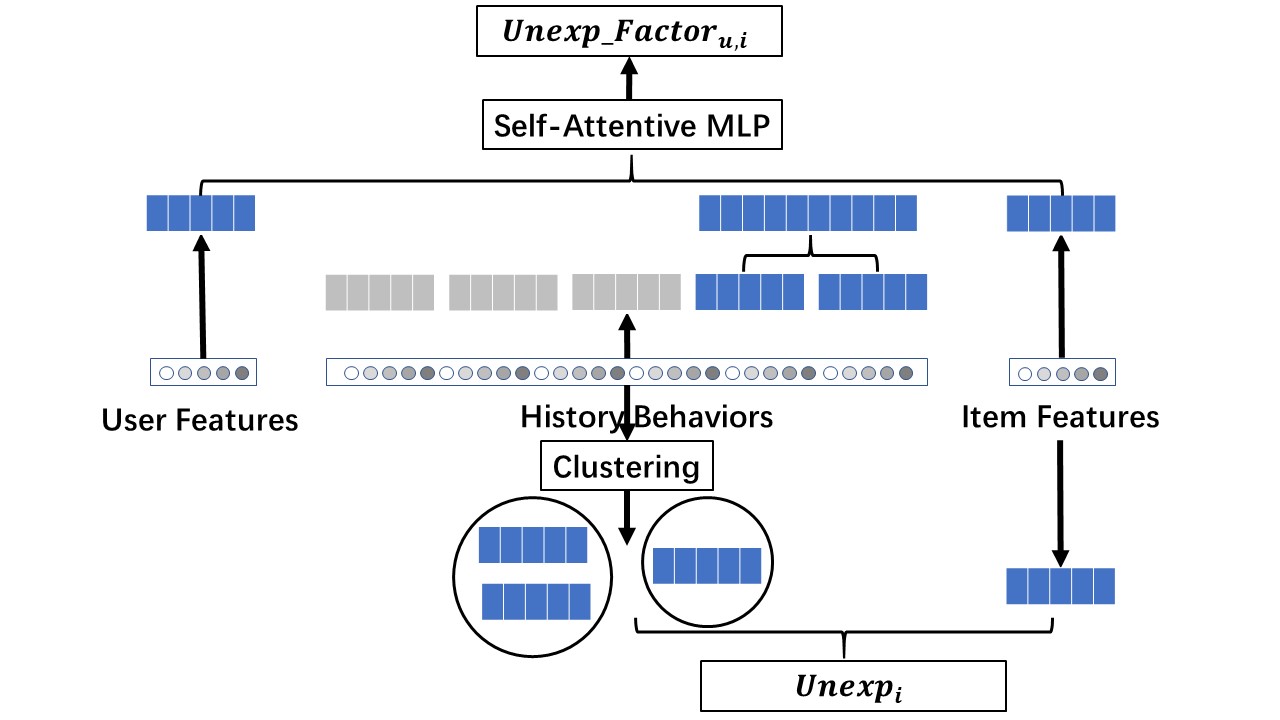}
        \caption{Unexpected}
    \end{subfigure}%
\caption{Overview of the proposed PURS model. The base model estimates the click-through-rate of certain user-item pairs, while the unexpected model captures unexpectedness of the new recommendation as well as user perception towards unexpectedness.}
\label{model}
\end{figure*}

The proposed recommendation model is presented in Figure \ref{model}, which consists of two components: the base model, which estimates the click-through-rate of certain user-item pairs, as we will discuss in this section; and the unexpected model, which captures unexpectedness of the new recommendation as well as user perception towards unexpectedness, as we have discussed in the last section. The unexpected recommendations are provided through joint optimization of the utility function.

\subsection{User and Item Embeddings}
To effectively identify user interests and provide corresponding item recommendations, it is crucial to capture the feature-level information about the users and the items, which can be used for recommendation purposes in many different ways.

In addition, the intrinsic nature of users and items play an important role in determining the success of the recommendations. Some people, for example heavy users of the online video platform, tend to trust more on the recommendations provided by the platform. Thus for a certain recommendation for them, the utility function tend to be higher. On the other hand, the quality of an item is also crucial for the click-through-rate estimation, for the better quality of recommended items would lead to an increase of user satisfaction. 

In this paper, we capture this user and item information in the form of \textit{embeddings} in the latent space and utilize the deep-learning based autoencoder approach to obtain these user and item embeddings and also to capture their interactions in the latent space. Specifically, we denote the explicit features for user $u$ and item $i$ as $W_{u} = [w_{u_{1}}, w_{u_{2}}, \cdots, w_{u_{m}}]$ and $W_{i} = [w_{i_{1}}, w_{i_{2}}, \cdots, w_{i_{n}}]$ respectively, where $m$ and $n$ stand for the dimensionality of user and item feature vectors. The goal is to train two separate neural networks: the encoder network that maps feature vectors into latent embeddings, and the decoder network that reconstructs feature vectors from latent embeddings. Due to effectiveness and efficiency of the training process, we represent both the encoder and the decoder as multi-layer perceptron (MLP). The MLP network learns the hidden representations by optimization reconstruction loss $L$:
\begin{equation}
L = ||W_{u}-MLP_{dec}^{u}(MLP_{enc}^{u}(W_{u}))||
\end{equation}
\begin{equation}
L = ||W_{i}-MLP_{dec}^{i}(MLP_{enc}^{i}(W_{i}))||
\end{equation}
where $MLP_{enc}$ and $MLP_{dec}$ represents the MLP network for encoder and decoder respectively. The MLP network is separately trained for obtaining user embedding and item embeddings, and is updated through back-propagation from the utility function optimization.

\subsection{Click-Through-Rate Estimation ($r_{u,i}$) using Self-Attentive GRU}
For a specific recommended item $i$, our goal is to predict whether user $u$ would click on this recommendation or not, which largely depends on the matching between the content of item $i$ and the interest of user $u$. This indicates the importance of precisely inferring user preferences from historic behaviors.

We denote the previous consumption of user $u$ as the sequence $P_{u} = [i_{u_{1}}, i_{u_{2}}, \cdots, i_{u_{K}}]$ The click-through-rate prediction model obtains a fixed-length representation vector of user interests by pooling all the embedding vectors over the user behavior sequence $P_{u}$. We follow the idea of sequence modeling and utilize the bidirectional GRU \cite{cho2014learning} neural networks to obtain the sequence embeddings. It is important to use the recurrent neural network to model user interests, for it is capable of capturing the time information and the order of user purchase, as more recent behavior would naturally have a higher impact on the current recommendation than previous actions. In addition, compared with other recurrent models like RNN or LSTM, GRU is computationally more efficient and better extracts semantic relationships \cite{chung2014empirical}. 

During the training process, we first map the behavior sequence to the corresponding item embeddings obtained in the previous stage. To illustrate the GRU learning procedure, we denote $W_{z}$,$W_{r}$,$U_{z}$ and $U_{r}$ as the weight matrices of current information and the past information for the update gate and the reset gate respectively. $x_{t}$ is the behavior embedding input at timestep $t$,  $h_{t}$ stands for the output user interest vector, $z_{t}$ denotes the update gate status and $r_{t}$ represents the status of reset gate. The hidden state at timestep $t$ could be obtained following these equations:
\begin{equation}
z_t = \sigma_g(W_{z} x_t + U_{z} h_{t-1} + b_z) 
\end{equation}
\begin{equation}
r_t = \sigma_g(W_{r} x_t + U_{r} h_{t-1} + b_r) 
\end{equation}
\begin{equation}
h_t =  (1-z_t) \circ h_{t-1} + z_t \circ \sigma_h(W_{h} x_t + U_{h} (r_t \circ h_{t-1}) + b_h)
\end{equation}

Note that, each historic consumption might have different influence on the current recommendation. For example, if a user have watched the documentary ``Deep Blue'' from BBC and is very satisfied with the viewing experience, the user will be more likely to accept the current recommendation of documentary ``Earth'', also from BBC. Meanwhile, the historic record of watching James Bond might have relatively smaller influence towards the acceptance of that documentary recommendation. Therefore, to capture the item-level heterogeneity in the behavior sequence, we propose to incorporate self-attentive mechanism \cite{shaw2018self} during the sequence modeling process. Typically,  each output element $s_{t}$ is computed as weighted sum of a linearly transformed input elements 
\begin{equation}
s_{t}=\sum_{i=1}^{n}\alpha_{ti}(x_{i}W^{t})
\end{equation}
Each weight coefficient $\alpha_{ti}$ is computed using a softmax function 
\begin{equation}
\alpha_{ti}=\frac{\exp{e_{ti}}}{\sum_{i=1}^{n}\exp{e_{ti}}}
\end{equation}
where $e_{ti}$ is computed using a compatibility function that compares two input elements $x_{t}$ and $x_{i}$ correspondingly.

By iteratively calculating hidden step throughout every time step, we obtain final hidden state at the end of the behavior sequence, which constitutes the user interest embeddings $R_{u}$ that captures the latent semantic information of the user's historic actions. To provide the click-through-rate estimation, we concatenate the obtained user interest embeddings $R_{u}$ with user embeddings $E_{u}$ and item embeddings $E_{i}$ extracted in the previous section and feed into a MLP network to get the prediction: $r_{u,i} = MLP(R_{u};E_{u};E_{i})$

\subsection{Unexpected Factor ($unexp\_factor_{u,i}$) using Self-Attentive MLP}
As we have discussed in the previous section, different people might have difference preferences towards unexpected recommendations, and their perception of unexpectedness is also influenced by the session-based information. Therefore, we need to take the user's historic actions in account when computing $unexp\_factor_{u,i}$ and provide personalized session-based recommendations.

We denote the previous consumption of user $u$ as the sequence $P_{u} = [i_{u_{1}}, i_{u_{2}}, \cdots, i_{u_{K}}]$ Following the idea of session-based recommendation \cite{ludewig2018evaluation}, instead of using the entire sequence to measure the factor of unexpectedness, we propose to only use a window of purchased items to identify the factor. The specific length of the window $K$ is a hyperparameter that we can adjust to get the optimal performance.

The most recent user actions in the window will then be extracted and serve as the input to the MLP network. To capture the heterogeneity of the extracted historic behavior towards current unexpected recommendations, we utilize the structure of local activation unit \cite{zhou2018deep} to determine whether the embedding of each item will be fed into the network. Instead of expressing the user's diverse interests with the same network structure, local activation unit is capable of adaptively calculating the relevance of historical behaviors towards current candidate recommendations.

Specifically, local activation unit performs a weighted sum pooling to adaptively calculate the activation stage of each behavior embedding and generate one single representation. We denote the sequence of item embeddings for user $u$ in the session-window as $P_{u} = [E_{i_{1}}, E_{i_{2}}, \cdots, E_{i_{K}}]$ For user $u$ and item $i$, the unexpected factor $unexp\_factor_{u,i}$ for this user-item pair will be calculated as
\begin{equation}
unexp\_factor_{u,i} = MLP(E_{u};\sum_{j=1}^{K}a(E_{u}, E_{i_{j}}, E_{i})E_{i_{j}};E_{i})
\end{equation}
where $a(\cdot)$ is a feed-forward network with output as the activation weight for each past purchase.

\section{Experiments}
In this section, we introduce extensive experiments that validate the superior recommendation performance of the proposed model in terms of both accuracy and novelty measures. The hyperparameters are optimized through Bayesian optimization \cite{snoek2012practical}. For all experiments, we select $K=10$ and use SGD as the optimizer with learning rate 1 and exponential decay 0.1. The dimensionality of embedding vector is 32. Layers of MLP is set by $32\times 64\times 1$. The batch size is set as 32. The codes are publicly available\footnote{Codes are available at \url{https://github.com/lpworld/PURS}}.

\subsection{Data}
We implement our model on three real-world datasets: the Yelp Challenge Dataset \footnote{https://www.yelp.com/dataset/challenge}, which contains check-in information of users and restaurants; the MovieLens Dataset \footnote{https://grouplens.org/datasets/movielens/}, which contains informations of user, movies and ratings; and the Youku dataset collected from the major online video platform Youku, which contains rich information of users, videos, clicks and their corresponding features. We list the descriptive statistics of these datasets in Table \ref{statisticalnumber}. For the click-through-rate prediction purposes, we binarize the ratings in Yelp and MovieLens datasets using the threshold 3.5 and transfer them into labels of click and non-click.

\begin{table}[h]
\centering
\begin{tabular}{|c|c|c|c|}
\hline
Dataset & \textbf{Yelp} & \textbf{MovieLens} & \textbf{Youku}\\ \hline
\# of Ratings & 2,254,589 & 19,961,113 & 1,806,157 \\ \hline
\# of Users & 76,564 & 138,493 & 46,143 \\ \hline
\# of Items & 75,231 & 15,079 & 53,657 \\ \hline
Sparsity & 0.039\% & 0.956\% & 0.073\% \\ \hline
\end{tabular}
\caption{Descriptive Statistics of Three Datasets}
\label{statisticalnumber}
\end{table}

\subsection{Baselines and Evaluation Metrics}
To illustrate that the proposed model indeed provide unexpected and useful recommendations at the same time, we select two groups of state-of-the-art baselines for comparison: click-through-rate prediction models and unexpected recommendation models. The first category includes:
\begin{itemize}
\item \textbf{DIN \cite{zhou2018deep}} Deep Interest Network designs a local activation unit to adaptively learn the representation of user interests from historical behaviors with respect to a certain item.
\item \textbf{DeepFM \cite{guo2017deepfm}} DeepFM combines the power of factorization machines for recommendation and deep learning for feature learning in a new neural network architecture.
\item \textbf{Wide \& Deep \cite{cheng2016wide}} Wide \& Deep utilizes the wide model to handle the manually designed cross product features, and the deep model to extract nonlinear relations among features.
\item \textbf{PNN \cite{qu2016product}} Product-based Nerual Network model introduces an additional product layer to serve as the feature extractor.
\end{itemize}

The second baseline category includes:
\begin{itemize}
\item \textbf{SPR \cite{lu2012serendipitous}} Serendipitous Personalized Ranking extends traditional personalized ranking methods by considering item popularity in AUC optimization.
\item \textbf{Auralist \cite{zhang2012auralist}} Auralist is a personalized recommendation system that balances between the desired goals of accuracy, diversity, novelty and serendipity simultaneously.
\item \textbf{DPP \cite{chen2018fast}} The Determinantal Point Process utilizes a fast greedy MAP inference approach to generate relevant and diverse recommendations.
\item \textbf{HOM-LIN \cite{adamopoulos2015unexpectedness}} HOM-LIN is the state-of-the-art unexpected recommendation algorithm, which provides recommendations through the hybrid utility function. 
\end{itemize}

In addition, we select the following popular accuracy and novelty metrics for the evaluation process: \textbf{AUC}, which measures the goodness of recommendation order by ranking all the items with predicted CTR and comparing with the click information; \textbf{HR@10}, which measures the number of clicks in top 10 recommendations; \textbf{Unexpectedness}, which measures the recommendations to users of those items that are not included in their consideration sets and depart from what they would expect from the recommender system and is calculated as Equation (3); and \textbf{Coverage}, which measures as the percentage of distinct items in the recommendation over all distinct items in the dataset.

\section{Results}
\subsection{Recommendation Performance}
\begin{figure*}[!]
    \centering
    \begin{subfigure}[t]{0.33\textwidth}
        \centering
        \includegraphics[width=\textwidth]{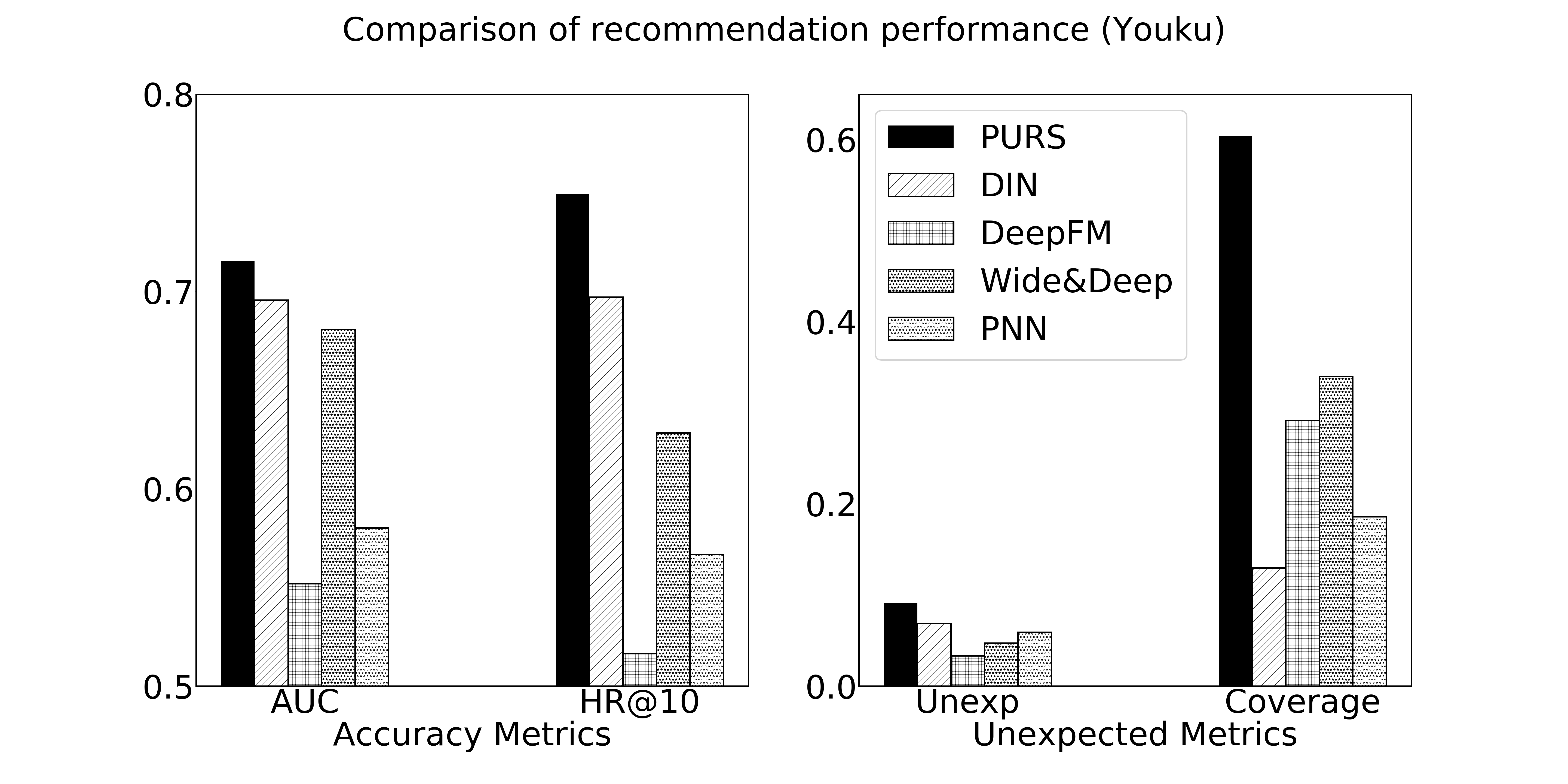}
        \caption{Youku Dataset}
    \end{subfigure}%
    ~ 
    \begin{subfigure}[t]{0.33\textwidth}
        \centering
        \includegraphics[width=\textwidth]{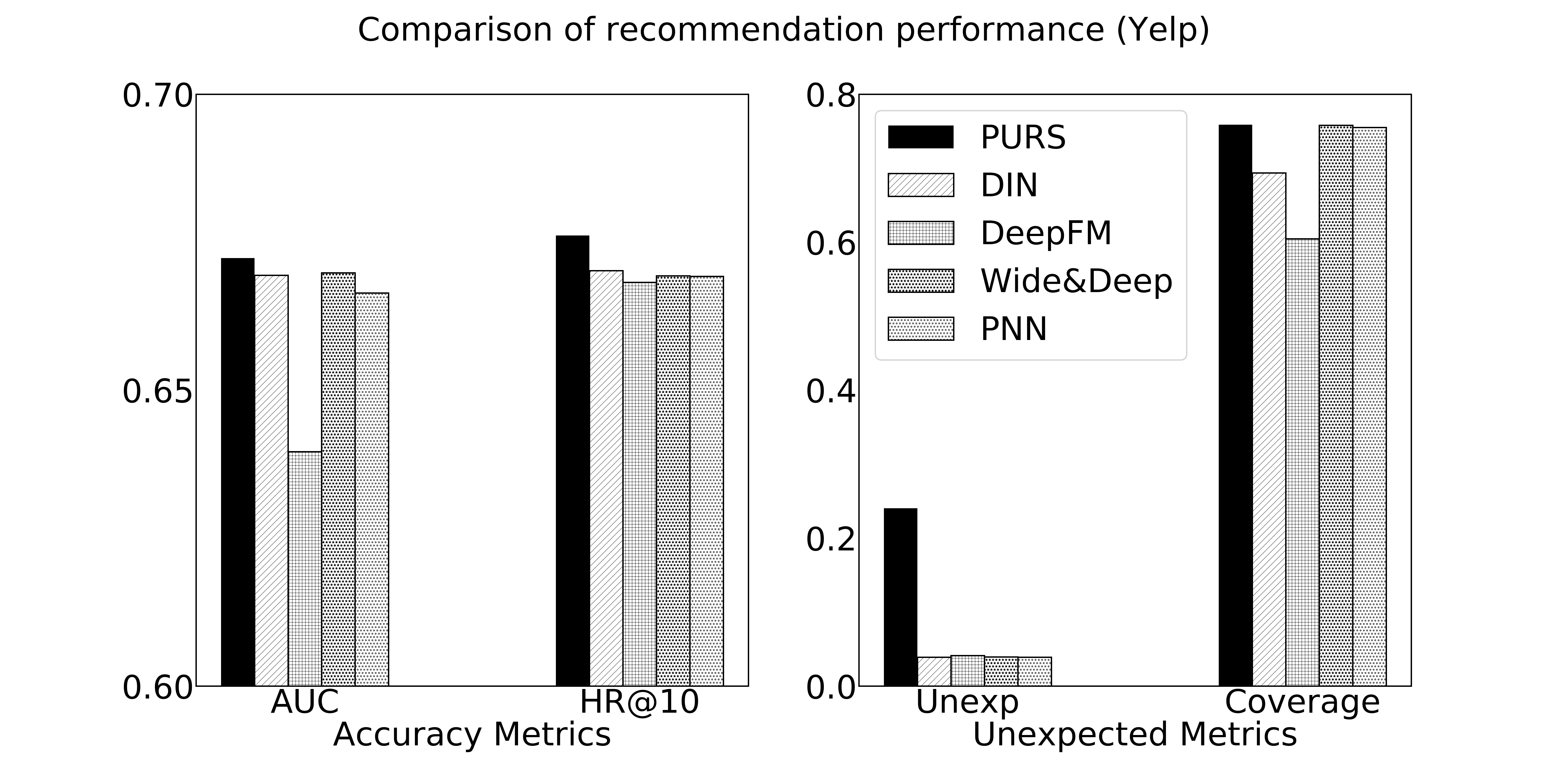}
        \caption{Yelp Dataset}
    \end{subfigure}%
    ~ 
    \begin{subfigure}[t]{0.33\textwidth}
        \centering
        \includegraphics[width=\textwidth]{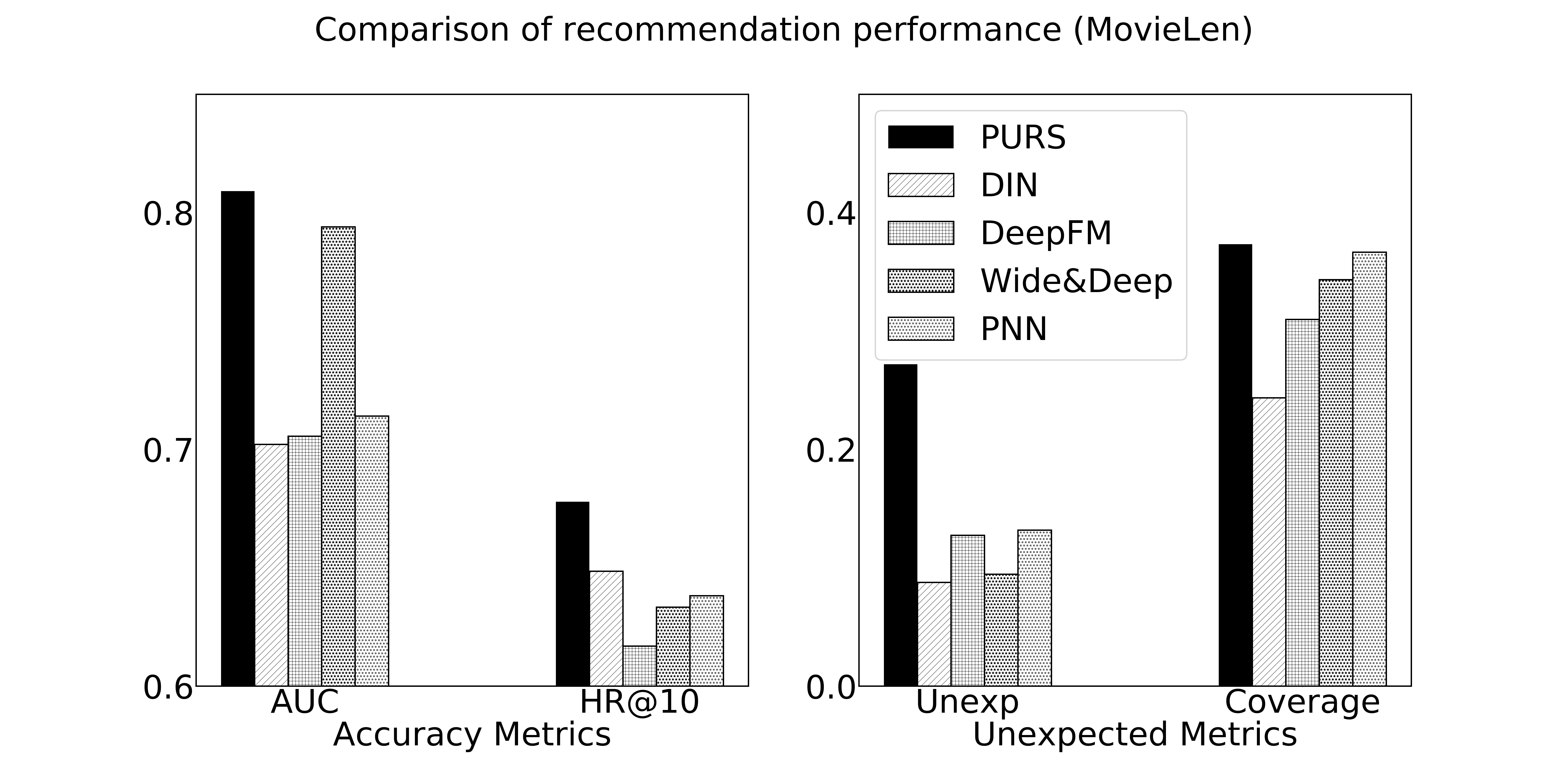}
        \caption{MovieLen Dataset}
    \end{subfigure}%
\caption{Comparison of recommendation performance in terms of accuracy and unexpectedness measures in three datasets.}
\label{comparison}
\end{figure*}

We implement the proposed PURS model and baseline methods in three real-world datasets. We conduct the time-stratified cross-validation with different initializations and report the average results over these experiments. As shown in Table \ref{unexp} and Figure \ref{comparison}, our proposed model consistently and significantly outperforms all other baselines in terms of both accuracy and unexpectedness measures across three datasets. Compare to the second-best baseline approach, we witness an increase of 2.75\% in AUC, 6.97\% in HR@10, 24.64\% in Unexpectedness and 43.71\% in Coverage measures. Especially in Youku Dataset where rich user behavior sequences in real business setting are available, PURS achieves the most significant improvement over other models. We also observe that all deep-learning based approaches performs significantly better than feature-based methods, which demonstrates the effectiveness of latent models. We conclude that our proposed approach achieves state-of-the-art unexpected recommendation performance and indeed provide satisfying and novel recommendations simultaneously to the target user.

\begin{table*}
\centering
\resizebox{\textwidth}{!}{
\begin{tabular}{|c|cccc|cccc|cccc|} \hline
\multirow{2}{*}{Algorithm} & \multicolumn{4}{c|}{Youku} & \multicolumn{4}{c|}{Yelp} & \multicolumn{4}{c|}{MovieLens} \\ \cline{2-13}
               &  AUC & HR@10 & Unexp & Coverage & AUC & HR@10 & Unexp & Coverage & AUC & HR@10 & Unexp & Coverage \\ \hline
\textbf{PURS} & \textbf{0.7154*} & \textbf{0.7494*} & \textbf{0.0913*} & \textbf{0.6040*} & \textbf{0.6723*} & \textbf{0.6761*} & \textbf{0.2401*} & \textbf{0.7585} & \textbf{0.8090*} & \textbf{0.6778*} & \textbf{0.2719*} & \textbf{0.3732*} \\ \hline
DIN & 0.6957 & 0.6972 & 0.0688 & 0.1298 & 0.6694 & 0.6702 & 0.0391 & 0.6934 & 0.7021 & 0.6485 & 0.0887 & 0.2435 \\ 
DeepFM & 0.5519 & 0.5164 & 0.0333 & 0.2919 & 0.6396 & 0.6682 & 0.0412 & 0.6044 & 0.7056 & 0.6169 & 0.1275 & 0.3098 \\ 
Wide\&Deep & 0.6807 & 0.6293 & 0.0472 & 0.3400 & 0.6698 & 0.6693 & 0.0392 & 0.7580 & 0.7940 & 0.6333 & 0.0944 & 0.3432 \\
PNN & 0.5801 & 0.5667 & 0.0593 & 0.1860 & 0.6664 & 0.6692 & 0.0391 & 0.7548 & 0.7140 & 0.6382 & 0.1318 & 0.3665 \\ \hline
HOM-LIN & 0.5812 & 0.5493 & 0.0602 & 0.4284 & 0.6287 & 0.6490 & 0.1433 & 0.5572 & 0.7177 & 0.5894 & 0.1116 & 0.1525 \\ 
Auralist & 0.5319 & 0.5250 & 0.0598 & 0.3990 & 0.6428 & 0.6104 & 0.1434 & 0.5442 & 0.6988 & 0.5710 & 0.1010 & 0.1333 \\
SPR & 0.5816 & 0.5156 & 0.0739 & 0.4668 & 0.6364 & 0.6492 & 0.1438 & 0.5849 & 0.7059 & 0.6122 & 0.1396 & 0.1728 \\ 
DPP & 0.6827 & 0.5777 & 0.0710 & 0.4702 & 0.5940 & 0.6414 & 0.1330 & 0.5072 & 0.7062 & 0.5984 & 0.1602 & 0.1908 \\ \hline
\end{tabular}
}
\caption{Comparison of recommendation performance in three datasets. The first block contains baselines for click-through-rate optimization, while the second block contains baselines for unexpectedness optimization. `*'  represents statistical significance at the 0.95 level.}
\label{unexp}
\end{table*}

\subsection{Ablation Study}
As discussed in the previous section, PURS achieves significant improvements over other baselines. These improvements indeed come from incorporating the following four components into the design of recommendation model: \textbf{Unexpectedness}, which aims at providing novel and satisfying recommendations; \textbf{Unexpected Activation Function}, which adjusts the input of unexpectedness into the utility function; \textbf{Personalized and Session-Based Factor}, which captures the user and session-level heterogeneity of perception towards unexpectedness; and \textbf{Clustering of Behavior Sequence}, which extracts the diverse user interests and constructs user expectations.

In this section, we conduct the ablation study to justify the importance of each factor. Specifically, we compare the proposed model with the following variations:
\begin{itemize}
\item \textbf{PURS-Variation 1 (Gaussian Activation)} This model users Gaussian distribution to serve as the unexpected activation function in the original design, and provides recommendations based on the following utility function $Utility_{u,i} = b_{u} + b_{i} + r_{u,i} + exp(-unexp_{u,i}^2)*unexp\_factor_{u,i}$
\item \textbf{PURS-Variation 2 (No Activation)} This model removes the unexpected activation function in the original design, and provides recommendations based on the following utility function $Utility_{u,i} = b_{u} + b_{i} + r_{u,i} + unexp_{u,i}*unexp\_factor_{u,i}$
\item \textbf{PURS-Variation 3 (No Unexpectedness Factor)} This model removes the unexpected activation function in the original design, and provides recommendations based on the following utility function $Utility_{u,i} = b_{u} + b_{i} + r_{u,i} + f(unexp_{u,i})$
\item \textbf{PURS-Variation 4 (No Unexpectedness)} This model removes the unexpected activation function in the original design, and provides recommendations based on the following utility function $Utility_{u,i} = b_{u} + b_{i} + r_{u,i}$
\item \textbf{PURS-Variation 5 (Single Closure of User Interest)} This model provides the unexpected recommendations using the same utility function, but instead remove the clustering procedure and model unexpectedness following \cite{li2019latent3} as $unexp_{i_{*},u} = d(w_{*}, C_{u})$ where $C_{u}$ is the entire latent closure generated by all past transactions of user $u$.
\end{itemize}

As shown in Table \ref{ablation}, if we remove any of these four components out of the recommendation model, we will witness significant loss in both accuracy and unexpectedness measures, especially in coverage metric. Therefore, the ablation study demonstrates that the superiority of our proposed model really comes from the combination of four novel components that all play significant role in contributing to satisfying and unexpected recommendations.

\begin{table*}
\centering
\resizebox{\textwidth}{!}{
\begin{tabular}{|c|cccc|cccc|cccc|} \hline
\multirow{2}{*}{Algorithm} & \multicolumn{4}{c|}{Youku} & \multicolumn{4}{c|}{Yelp} & \multicolumn{4}{c|}{MovieLens} \\ \cline{2-13}
               &  AUC & HR@10 & Unexp & Coverage & AUC & HR@10 & Unexp & Coverage & AUC & HR@10 & Unexp & Coverage \\ \hline
\textbf{PURS} & \textbf{0.7154*} & \textbf{0.7494*} & \textbf{0.0913*} & \textbf{0.6040*} & \textbf{0.6723*} & \textbf{0.6761*} & \textbf{0.2401*} & \textbf{0.7585*} & \textbf{0.8090*} & \textbf{0.6778*} & \textbf{0.2719*} & \textbf{0.3732*} \\ \hline
PURS-Variation 1 & 0.6826 & 0.7292 & 0.0828 & 0.5548 & 0.6682 & 0.6602 & 0.1298 & 0.7292 & 0.7757 & 0.6419 & 0.1812 & 0.3350 \\
PURS-Variation 2 & 0.7067 & 0.7148 & 0.0707 & 0.3026 & 0.6692 & 0.6630 & 0.0412 & 0.7580 & 0.8041 & 0.6585 & 0.2471 & 0.3580 \\ 
PURS-Variation 3 & 0.7036 & 0.6720 & 0.0762 & 0.1522 & 0.6508 & 0.6692 & 0.0391 & 0.7583 & 0.7666 & 0.6460 & 0.0888 & 0.2792 \\ 
PURS-Variation 4 & 0.7063 & 0.5628 & 0.0688 & 0.1298 & 0.6702 & 0.6702 & 0.0391 & 0.6934 & 0.7586 & 0.6331 & 0.0887 & 0.2435 \\ 
PURS-Variation 5 & 0.7038 & 0.7080 & 0.0477 & 0.2042 & 0.6701 & 0.6700 & 0.0395 & 0.7580 & 0.7715 & 0.6596 & 0.1727 & 0.3561 \\ \hline
\end{tabular}
}
\caption{Ablation Study of recommendation performance in three datasets. `*'  represents statistical significance at the 0.95 level.}
\label{ablation}
\end{table*}

\subsection{Improving Accuracy and Novelty Simultaneously}
\begin{figure}
\centering
\includegraphics[width=0.5\textwidth]{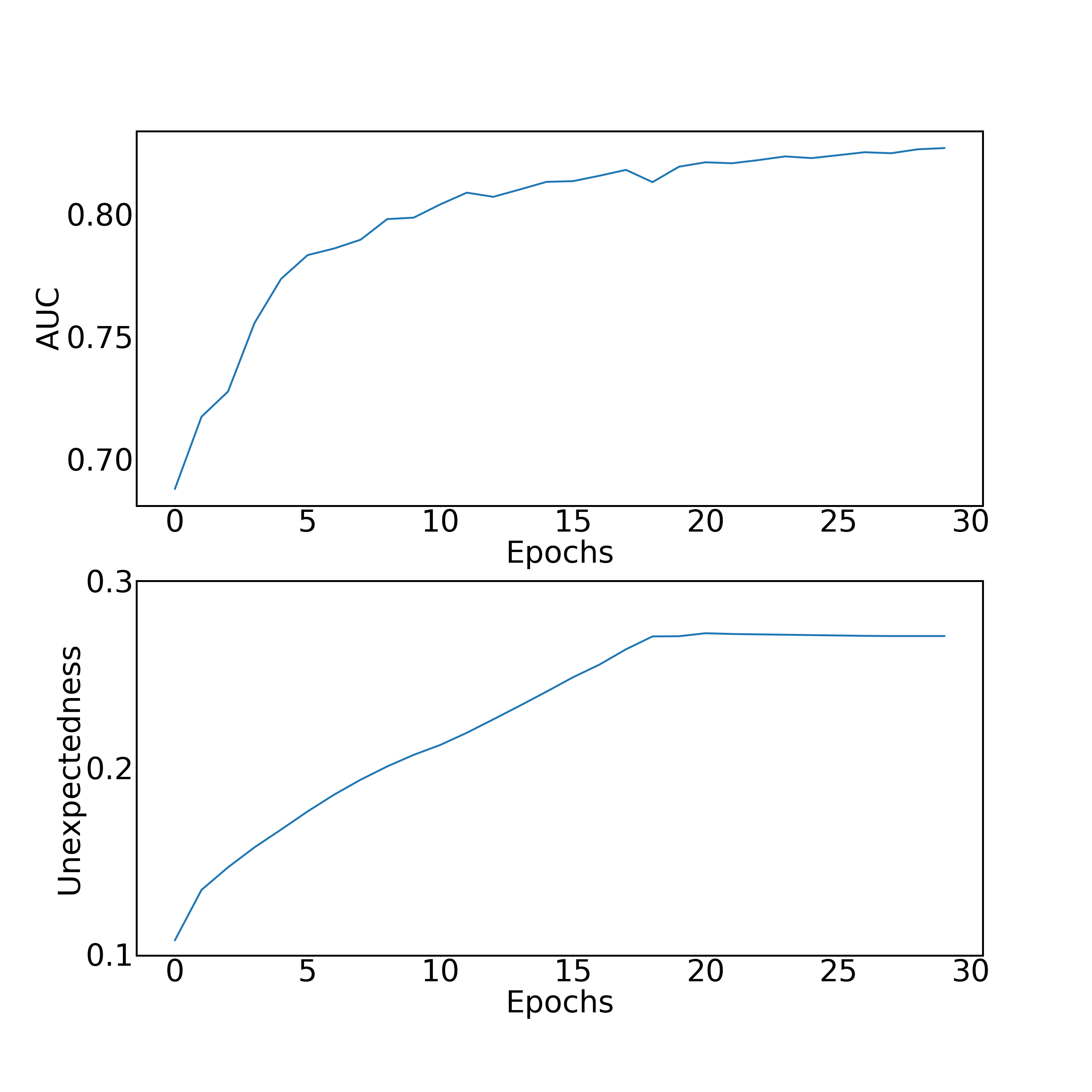}
\caption{Improving accuracy and unexpectedness measures simultaneously in each training epoch of PURS in the MovieLens dataset} 
\label{tradeoff}
\end{figure}

In Table \ref{unexp}, we observe that unexpectedness-oriented baselines generally achieve better performance in unexpected measures, but at the cost of losing accuracy measures, when comparing to the CTR-oriented baselines. This observation is in line with the prior literature \cite{adomavicius2008overcoming,zhou2010solving} discussing the trade-off between these two objectives. However, our proposed PURS model manages to surpass baselines models in both \textbf{accuracy} and \textbf{novelty} measures at the same time. In addition, PURS is capable of improving AUC and unexpectedness metrics simultaneously throughout the training process, as shown in Figure \ref{tradeoff}.

\subsection{Scalability}
\begin{figure}
\centering
\includegraphics[width=0.5\textwidth]{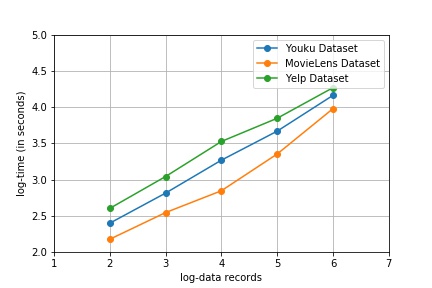}
\caption{Scalability of PURS on the Yelp, MovieLens and Youku datasets with increasing data sizes from 100 to 1,000,000 records.} 
\label{scalability}
\end{figure}

To test for scalability, we provide recommendations using PURS with aforementioned parameter values for the Yelp, MovieLens and Youku datasets with increasing data sizes from 100 to 1,000,000 records. As shown in Figure \ref{scalability}, we empirically observe that the proposed PURS model scales linearly with increase in number of data records to finish the training process and provide unexpected recommendations accordingly. The training procedure comprises of obtaining user and item embeddings and jointly optimizing the utility function. The optimization phase is made efficient using batch normalization \cite{ioffe2015batch} and distributed training. As our experiments confirm, PURS is capable of learning network parameters efficiently and indeed scales well.

\begin{table*}
\centering
\begin{tabular}{|c|cccccc|} \hline
 & VV & TS & ID & CTR  & Unexpectedness & Coverage \\ \hline
Improvement & +3.74\%* & +4.63\%* & +4.13\%* & +0.80\%* & +9.74\%* & +1.23\%* \\ \hline
\end{tabular}
\caption{Unexpected recommendation performance in online A/B test: performance increase compared to the current model. `*'  represents statistical significance at the 0.95 level.}
\label{online}
\end{table*}

\section{Online A/B Test}
We conduct the online A/B test at Alibaba-Youku, a major video recommendation platform from 2019-11 to 2019-12. During the testing period, we compare the proposed PURS model with the latest production model in the company. We measure the performance using standard business metrics: \textbf{VV} (Video View, average video viewed by each user), \textbf{TS} (Time Spent, average time that each user spends on the platform), \textbf{ID} (Impression Depth, average impression through one session) and \textbf{CTR} (Click-Through-Rate, the percentage of user clicking on the recommended video). We also measure the novelty of the recommended videos using the unexpectedness and coverage measures described in Section 5.2. We present the results in Table \ref{online} that demonstrates significant and consistent improvements over the current model in all the four business metrics, and these improvements indeed come from providing more unexpected recommendations, as demonstrated in Section 6. The proposed model is in the process of being deployed by the company.

\section{Conclusions}
In this paper, we present the PURS model that incorporates unexpectedness into the recommendation process. Specifically, we propose to model user interests as clusters of embedding closures in the latent space and calculate unexpectedness as the weighted distance between the new items and the interest clusters. We utilize the sequence modeling and the self-attention mechanism to capture personalized and session-based user perception of unexpectedness. We also propose a novel unexpected activation function to achieve better unexpected recommendation performance. We subsequently combine the CTR estimation with the degree of unexpectedness to provide final recommendations. Extensive offline and online experiments illustrate superiority of the proposed model.

As the future work, we plan to study the impact of unexpected recommendations on user behaviors. Also, we plan to model the user interests in a more explicit manner to provide better unexpected recommendations.

\balance
\bibliographystyle{ACM-Reference-Format}
\bibliography{sigproc}


\begin{thebibliography}{51}


\ifx \showCODEN    \undefined \def \showCODEN     #1{\unskip}     \fi
\ifx \showDOI      \undefined \def \showDOI       #1{#1}\fi
\ifx \showISBNx    \undefined \def \showISBNx     #1{\unskip}     \fi
\ifx \showISBNxiii \undefined \def \showISBNxiii  #1{\unskip}     \fi
\ifx \showISSN     \undefined \def \showISSN      #1{\unskip}     \fi
\ifx \showLCCN     \undefined \def \showLCCN      #1{\unskip}     \fi
\ifx \shownote     \undefined \def \shownote      #1{#1}          \fi
\ifx \showarticletitle \undefined \def \showarticletitle #1{#1}   \fi
\ifx \showURL      \undefined \def \showURL       {\relax}        \fi
\providecommand\bibfield[2]{#2}
\providecommand\bibinfo[2]{#2}
\providecommand\natexlab[1]{#1}
\providecommand\showeprint[2][]{arXiv:#2}

\bibitem[\protect\citeauthoryear{Adamopoulos}{Adamopoulos}{2014}]%
        {adamopoulos2014discovering}
\bibfield{author}{\bibinfo{person}{Panagiotis Adamopoulos}.}
  \bibinfo{year}{2014}\natexlab{}.
\newblock \showarticletitle{On discovering non-obvious recommendations: Using
  unexpectedness and neighborhood selection methods in collaborative filtering
  systems}. In \bibinfo{booktitle}{\emph{Proceedings of the 7th ACM
  international conference on Web search and data mining}}. ACM,
  \bibinfo{pages}{655--660}.
\newblock


\bibitem[\protect\citeauthoryear{Adamopoulos and Tuzhilin}{Adamopoulos and
  Tuzhilin}{2014}]%
        {adamopoulos2014over}
\bibfield{author}{\bibinfo{person}{Panagiotis Adamopoulos} {and}
  \bibinfo{person}{Alexander Tuzhilin}.} \bibinfo{year}{2014}\natexlab{}.
\newblock \showarticletitle{On over-specialization and concentration bias of
  recommendations: Probabilistic neighborhood selection in collaborative
  filtering systems}. In \bibinfo{booktitle}{\emph{Proceedings of the 8th ACM
  Conference on Recommender systems}}. ACM, \bibinfo{pages}{153--160}.
\newblock


\bibitem[\protect\citeauthoryear{Adamopoulos and Tuzhilin}{Adamopoulos and
  Tuzhilin}{2015}]%
        {adamopoulos2015unexpectedness}
\bibfield{author}{\bibinfo{person}{Panagiotis Adamopoulos} {and}
  \bibinfo{person}{Alexander Tuzhilin}.} \bibinfo{year}{2015}\natexlab{}.
\newblock \showarticletitle{On unexpectedness in recommender systems: Or how to
  better expect the unexpected}.
\newblock \bibinfo{journal}{\emph{ACM Transactions on Intelligent Systems and
  Technology (TIST)}} \bibinfo{volume}{5}, \bibinfo{number}{4}
  (\bibinfo{year}{2015}), \bibinfo{pages}{54}.
\newblock


\bibitem[\protect\citeauthoryear{Adomavicius and Kwon}{Adomavicius and
  Kwon}{[n.d.]}]%
        {adomavicius2008overcoming}
\bibfield{author}{\bibinfo{person}{Gediminas Adomavicius} {and}
  \bibinfo{person}{YoungOk Kwon}.} \bibinfo{year}{[n.d.]}\natexlab{}.
\newblock \showarticletitle{Overcoming accuracy-diversity tradeoff in
  recommender systems: a variance-based approach}. Citeseer.
\newblock


\bibitem[\protect\citeauthoryear{Buldygin and Kozachenko}{Buldygin and
  Kozachenko}{1980}]%
        {buldygin1980sub}
\bibfield{author}{\bibinfo{person}{Valerii~V Buldygin} {and}
  \bibinfo{person}{Yu~V Kozachenko}.} \bibinfo{year}{1980}\natexlab{}.
\newblock \showarticletitle{Sub-Gaussian random variables}.
\newblock \bibinfo{journal}{\emph{Ukrainian Mathematical Journal}}
  \bibinfo{volume}{32}, \bibinfo{number}{6} (\bibinfo{year}{1980}),
  \bibinfo{pages}{483--489}.
\newblock


\bibitem[\protect\citeauthoryear{Chen, Yang, Wang, Yang, and Yuan}{Chen
  et~al\mbox{.}}{2019}]%
        {chen2019serendipity}
\bibfield{author}{\bibinfo{person}{Li Chen}, \bibinfo{person}{Yonghua Yang},
  \bibinfo{person}{Ningxia Wang}, \bibinfo{person}{Keping Yang}, {and}
  \bibinfo{person}{Quan Yuan}.} \bibinfo{year}{2019}\natexlab{}.
\newblock \showarticletitle{How Serendipity Improves User Satisfaction with
  Recommendations? A Large-Scale User Evaluation}. In
  \bibinfo{booktitle}{\emph{The World Wide Web Conference}}. ACM,
  \bibinfo{pages}{240--250}.
\newblock


\bibitem[\protect\citeauthoryear{Chen, Zhang, and Zhou}{Chen
  et~al\mbox{.}}{2018}]%
        {chen2018fast}
\bibfield{author}{\bibinfo{person}{Laming Chen}, \bibinfo{person}{Guoxin
  Zhang}, {and} \bibinfo{person}{Eric Zhou}.} \bibinfo{year}{2018}\natexlab{}.
\newblock \showarticletitle{Fast greedy MAP inference for Determinantal Point
  Process to improve recommendation diversity}. In
  \bibinfo{booktitle}{\emph{Advances in Neural Information Processing
  Systems}}. \bibinfo{pages}{5622--5633}.
\newblock


\bibitem[\protect\citeauthoryear{Cheng, Koc, Harmsen, Shaked, Chandra, Aradhye,
  Anderson, Corrado, Chai, Ispir, et~al\mbox{.}}{Cheng et~al\mbox{.}}{2016}]%
        {cheng2016wide}
\bibfield{author}{\bibinfo{person}{Heng-Tze Cheng}, \bibinfo{person}{Levent
  Koc}, \bibinfo{person}{Jeremiah Harmsen}, \bibinfo{person}{Tal Shaked},
  \bibinfo{person}{Tushar Chandra}, \bibinfo{person}{Hrishi Aradhye},
  \bibinfo{person}{Glen Anderson}, \bibinfo{person}{Greg Corrado},
  \bibinfo{person}{Wei Chai}, \bibinfo{person}{Mustafa Ispir}, {et~al\mbox{.}}}
  \bibinfo{year}{2016}\natexlab{}.
\newblock \showarticletitle{Wide \& deep learning for recommender systems}. In
  \bibinfo{booktitle}{\emph{Proceedings of the 1st workshop on deep learning
  for recommender systems}}. ACM, \bibinfo{pages}{7--10}.
\newblock


\bibitem[\protect\citeauthoryear{Cheng}{Cheng}{1995}]%
        {cheng1995mean}
\bibfield{author}{\bibinfo{person}{Yizong Cheng}.}
  \bibinfo{year}{1995}\natexlab{}.
\newblock \showarticletitle{Mean shift, mode seeking, and clustering}.
\newblock \bibinfo{journal}{\emph{IEEE transactions on pattern analysis and
  machine intelligence}} \bibinfo{volume}{17}, \bibinfo{number}{8}
  (\bibinfo{year}{1995}), \bibinfo{pages}{790--799}.
\newblock


\bibitem[\protect\citeauthoryear{Cho, Van~Merri{\"e}nboer, Gulcehre, Bahdanau,
  Bougares, Schwenk, and Bengio}{Cho et~al\mbox{.}}{2014}]%
        {cho2014learning}
\bibfield{author}{\bibinfo{person}{Kyunghyun Cho}, \bibinfo{person}{Bart
  Van~Merri{\"e}nboer}, \bibinfo{person}{Caglar Gulcehre},
  \bibinfo{person}{Dzmitry Bahdanau}, \bibinfo{person}{Fethi Bougares},
  \bibinfo{person}{Holger Schwenk}, {and} \bibinfo{person}{Yoshua Bengio}.}
  \bibinfo{year}{2014}\natexlab{}.
\newblock \showarticletitle{Learning phrase representations using RNN
  encoder-decoder for statistical machine translation}.
\newblock \bibinfo{journal}{\emph{arXiv preprint arXiv:1406.1078}}
  (\bibinfo{year}{2014}).
\newblock


\bibitem[\protect\citeauthoryear{Chung, Gulcehre, Cho, and Bengio}{Chung
  et~al\mbox{.}}{2014}]%
        {chung2014empirical}
\bibfield{author}{\bibinfo{person}{Junyoung Chung}, \bibinfo{person}{Caglar
  Gulcehre}, \bibinfo{person}{KyungHyun Cho}, {and} \bibinfo{person}{Yoshua
  Bengio}.} \bibinfo{year}{2014}\natexlab{}.
\newblock \showarticletitle{Empirical evaluation of gated recurrent neural
  networks on sequence modeling}.
\newblock \bibinfo{journal}{\emph{arXiv preprint arXiv:1412.3555}}
  (\bibinfo{year}{2014}).
\newblock


\bibitem[\protect\citeauthoryear{Comaniciu and Meer}{Comaniciu and
  Meer}{2002}]%
        {comaniciu2002mean}
\bibfield{author}{\bibinfo{person}{Dorin Comaniciu} {and}
  \bibinfo{person}{Peter Meer}.} \bibinfo{year}{2002}\natexlab{}.
\newblock \showarticletitle{Mean shift: A robust approach toward feature space
  analysis}.
\newblock \bibinfo{journal}{\emph{IEEE Transactions on Pattern Analysis \&
  Machine Intelligence}} \bibinfo{number}{5} (\bibinfo{year}{2002}),
  \bibinfo{pages}{603--619}.
\newblock


\bibitem[\protect\citeauthoryear{Davis}{Davis}{1959}]%
        {davis1959leonhard}
\bibfield{author}{\bibinfo{person}{Philip~J Davis}.}
  \bibinfo{year}{1959}\natexlab{}.
\newblock \showarticletitle{Leonhard Euler's integral: A historical profile of
  the Gamma function: In memoriam: Milton Abramowitz}.
\newblock \bibinfo{journal}{\emph{The American Mathematical Monthly}}
  \bibinfo{volume}{66}, \bibinfo{number}{10} (\bibinfo{year}{1959}),
  \bibinfo{pages}{849--869}.
\newblock


\bibitem[\protect\citeauthoryear{Dong, Chawla, and Swami}{Dong
  et~al\mbox{.}}{2017}]%
        {dong2017metapath2vec}
\bibfield{author}{\bibinfo{person}{Yuxiao Dong}, \bibinfo{person}{Nitesh~V
  Chawla}, {and} \bibinfo{person}{Ananthram Swami}.}
  \bibinfo{year}{2017}\natexlab{}.
\newblock \showarticletitle{metapath2vec: Scalable representation learning for
  heterogeneous networks}. In \bibinfo{booktitle}{\emph{Proceedings of the 23rd
  ACM SIGKDD International Conference on Knowledge Discovery and Data Mining}}.
  ACM, \bibinfo{pages}{135--144}.
\newblock


\bibitem[\protect\citeauthoryear{Feng, Lv, Shen, Wang, Sun, Zhu, and Yang}{Feng
  et~al\mbox{.}}{2019}]%
        {feng2019deep}
\bibfield{author}{\bibinfo{person}{Yufei Feng}, \bibinfo{person}{Fuyu Lv},
  \bibinfo{person}{Weichen Shen}, \bibinfo{person}{Menghan Wang},
  \bibinfo{person}{Fei Sun}, \bibinfo{person}{Yu Zhu}, {and}
  \bibinfo{person}{Keping Yang}.} \bibinfo{year}{2019}\natexlab{}.
\newblock \showarticletitle{Deep session interest network for click-through
  rate prediction}.
\newblock \bibinfo{journal}{\emph{arXiv preprint arXiv:1905.06482}}
  (\bibinfo{year}{2019}).
\newblock


\bibitem[\protect\citeauthoryear{Gartrell, Paquet, and Koenigstein}{Gartrell
  et~al\mbox{.}}{2017}]%
        {gartrell2017low}
\bibfield{author}{\bibinfo{person}{Mike Gartrell}, \bibinfo{person}{Ulrich
  Paquet}, {and} \bibinfo{person}{Noam Koenigstein}.}
  \bibinfo{year}{2017}\natexlab{}.
\newblock \showarticletitle{Low-rank factorization of determinantal point
  processes}. In \bibinfo{booktitle}{\emph{Thirty-First AAAI Conference on
  Artificial Intelligence}}.
\newblock


\bibitem[\protect\citeauthoryear{Ge, Delgado-Battenfeld, and Jannach}{Ge
  et~al\mbox{.}}{2010}]%
        {ge2010beyond}
\bibfield{author}{\bibinfo{person}{Mouzhi Ge}, \bibinfo{person}{Carla
  Delgado-Battenfeld}, {and} \bibinfo{person}{Dietmar Jannach}.}
  \bibinfo{year}{2010}\natexlab{}.
\newblock \showarticletitle{Beyond accuracy: evaluating recommender systems by
  coverage and serendipity}. In \bibinfo{booktitle}{\emph{Proceedings of the
  fourth ACM conference on Recommender systems}}. ACM,
  \bibinfo{pages}{257--260}.
\newblock


\bibitem[\protect\citeauthoryear{Guo, Tang, Ye, Li, and He}{Guo
  et~al\mbox{.}}{2017}]%
        {guo2017deepfm}
\bibfield{author}{\bibinfo{person}{Huifeng Guo}, \bibinfo{person}{Ruiming
  Tang}, \bibinfo{person}{Yunming Ye}, \bibinfo{person}{Zhenguo Li}, {and}
  \bibinfo{person}{Xiuqiang He}.} \bibinfo{year}{2017}\natexlab{}.
\newblock \showarticletitle{DeepFM: a factorization-machine based neural
  network for CTR prediction}.
\newblock \bibinfo{journal}{\emph{arXiv preprint arXiv:1703.04247}}
  (\bibinfo{year}{2017}).
\newblock


\bibitem[\protect\citeauthoryear{He, Liao, Zhang, Nie, Hu, and Chua}{He
  et~al\mbox{.}}{2017}]%
        {he2017neural}
\bibfield{author}{\bibinfo{person}{Xiangnan He}, \bibinfo{person}{Lizi Liao},
  \bibinfo{person}{Hanwang Zhang}, \bibinfo{person}{Liqiang Nie},
  \bibinfo{person}{Xia Hu}, {and} \bibinfo{person}{Tat-Seng Chua}.}
  \bibinfo{year}{2017}\natexlab{}.
\newblock \showarticletitle{Neural collaborative filtering}. In
  \bibinfo{booktitle}{\emph{Proceedings of the 26th international conference on
  world wide web}}. International World Wide Web Conferences Steering
  Committee, \bibinfo{pages}{173--182}.
\newblock


\bibitem[\protect\citeauthoryear{Herlocker, Konstan, Terveen, and
  Riedl}{Herlocker et~al\mbox{.}}{2004}]%
        {herlocker2004evaluating}
\bibfield{author}{\bibinfo{person}{Jonathan~L Herlocker},
  \bibinfo{person}{Joseph~A Konstan}, \bibinfo{person}{Loren~G Terveen}, {and}
  \bibinfo{person}{John~T Riedl}.} \bibinfo{year}{2004}\natexlab{}.
\newblock \showarticletitle{Evaluating collaborative filtering recommender
  systems}.
\newblock \bibinfo{journal}{\emph{ACM Transactions on Information Systems
  (TOIS)}} \bibinfo{volume}{22}, \bibinfo{number}{1} (\bibinfo{year}{2004}),
  \bibinfo{pages}{5--53}.
\newblock


\bibitem[\protect\citeauthoryear{Hinton and Salakhutdinov}{Hinton and
  Salakhutdinov}{2006}]%
        {hinton2006reducing}
\bibfield{author}{\bibinfo{person}{Geoffrey~E Hinton} {and}
  \bibinfo{person}{Ruslan~R Salakhutdinov}.} \bibinfo{year}{2006}\natexlab{}.
\newblock \showarticletitle{Reducing the dimensionality of data with neural
  networks}.
\newblock \bibinfo{journal}{\emph{science}} \bibinfo{volume}{313},
  \bibinfo{number}{5786} (\bibinfo{year}{2006}), \bibinfo{pages}{504--507}.
\newblock


\bibitem[\protect\citeauthoryear{Iaquinta, de~Gemmis, Lops, Semeraro, and
  Molino}{Iaquinta et~al\mbox{.}}{2010}]%
        {iaquinta2010can}
\bibfield{author}{\bibinfo{person}{Leo Iaquinta}, \bibinfo{person}{Marco de
  Gemmis}, \bibinfo{person}{Pasquale Lops}, \bibinfo{person}{Giovanni
  Semeraro}, {and} \bibinfo{person}{Piero Molino}.}
  \bibinfo{year}{2010}\natexlab{}.
\newblock \showarticletitle{Can a recommender system induce serendipitous
  encounters?}
\newblock In \bibinfo{booktitle}{\emph{E-commerce}}.
  \bibinfo{publisher}{InTech}.
\newblock


\bibitem[\protect\citeauthoryear{Ioffe and Szegedy}{Ioffe and Szegedy}{2015}]%
        {ioffe2015batch}
\bibfield{author}{\bibinfo{person}{Sergey Ioffe} {and}
  \bibinfo{person}{Christian Szegedy}.} \bibinfo{year}{2015}\natexlab{}.
\newblock \showarticletitle{Batch normalization: Accelerating deep network
  training by reducing internal covariate shift}. In
  \bibinfo{booktitle}{\emph{International conference on machine learning}}.
  \bibinfo{pages}{448--456}.
\newblock


\bibitem[\protect\citeauthoryear{Kapoor, Kumar, Terveen, Konstan, and
  Schrater}{Kapoor et~al\mbox{.}}{2015a}]%
        {kapoor2015like}
\bibfield{author}{\bibinfo{person}{Komal Kapoor}, \bibinfo{person}{Vikas
  Kumar}, \bibinfo{person}{Loren Terveen}, \bibinfo{person}{Joseph~A Konstan},
  {and} \bibinfo{person}{Paul Schrater}.} \bibinfo{year}{2015}\natexlab{a}.
\newblock \showarticletitle{I like to explore sometimes: Adapting to dynamic
  user novelty preferences}. In \bibinfo{booktitle}{\emph{Proceedings of the
  9th ACM Conference on Recommender Systems}}. ACM, \bibinfo{pages}{19--26}.
\newblock


\bibitem[\protect\citeauthoryear{Kapoor, Subbian, Srivastava, and
  Schrater}{Kapoor et~al\mbox{.}}{2015b}]%
        {kapoor2015just}
\bibfield{author}{\bibinfo{person}{Komal Kapoor}, \bibinfo{person}{Karthik
  Subbian}, \bibinfo{person}{Jaideep Srivastava}, {and} \bibinfo{person}{Paul
  Schrater}.} \bibinfo{year}{2015}\natexlab{b}.
\newblock \showarticletitle{Just in time recommendations: Modeling the dynamics
  of boredom in activity streams}. In \bibinfo{booktitle}{\emph{Proceedings of
  the Eighth ACM International Conference on Web Search and Data Mining}}. ACM,
  \bibinfo{pages}{233--242}.
\newblock


\bibitem[\protect\citeauthoryear{Li and Tuzhilin}{Li and Tuzhilin}{2019}]%
        {li2019latent3}
\bibfield{author}{\bibinfo{person}{Pan Li} {and} \bibinfo{person}{Alexander
  Tuzhilin}.} \bibinfo{year}{2019}\natexlab{}.
\newblock \showarticletitle{Latent Modeling of Unexpectedness for
  Recommendations}.
\newblock \bibinfo{journal}{\emph{Proceedings of ACM RecSys 2019 Late-breaking
  Results}} (\bibinfo{year}{2019}), \bibinfo{pages}{7--10}.
\newblock


\bibitem[\protect\citeauthoryear{Li and Tuzhilin}{Li and Tuzhilin}{2020}]%
        {li2020latent}
\bibfield{author}{\bibinfo{person}{Pan Li} {and} \bibinfo{person}{Alexander
  Tuzhilin}.} \bibinfo{year}{2020}\natexlab{}.
\newblock \showarticletitle{Latent Unexpected Recommendations}.
\newblock \bibinfo{journal}{\emph{Forthcoming at ACM Transactions on
  Intelligent Systems and Technology (TIST)}} (\bibinfo{year}{2020}).
\newblock


\bibitem[\protect\citeauthoryear{Lu, Chen, Zhang, Yang, and Yu}{Lu
  et~al\mbox{.}}{2012}]%
        {lu2012serendipitous}
\bibfield{author}{\bibinfo{person}{Qiuxia Lu}, \bibinfo{person}{Tianqi Chen},
  \bibinfo{person}{Weinan Zhang}, \bibinfo{person}{Diyi Yang}, {and}
  \bibinfo{person}{Yong Yu}.} \bibinfo{year}{2012}\natexlab{}.
\newblock \showarticletitle{Serendipitous personalized ranking for top-n
  recommendation}. In \bibinfo{booktitle}{\emph{Web Intelligence and
  Intelligent Agent Technology (WI-IAT), 2012 IEEE/WIC/ACM International
  Conferences on}}, Vol.~\bibinfo{volume}{1}. IEEE, \bibinfo{pages}{258--265}.
\newblock


\bibitem[\protect\citeauthoryear{Ludewig and Jannach}{Ludewig and
  Jannach}{2018}]%
        {ludewig2018evaluation}
\bibfield{author}{\bibinfo{person}{Malte Ludewig} {and}
  \bibinfo{person}{Dietmar Jannach}.} \bibinfo{year}{2018}\natexlab{}.
\newblock \showarticletitle{Evaluation of session-based recommendation
  algorithms}.
\newblock \bibinfo{journal}{\emph{User Modeling and User-Adapted Interaction}}
  \bibinfo{volume}{28}, \bibinfo{number}{4-5} (\bibinfo{year}{2018}),
  \bibinfo{pages}{331--390}.
\newblock


\bibitem[\protect\citeauthoryear{McAlister and Pessemier}{McAlister and
  Pessemier}{1982}]%
        {mcalister1982variety}
\bibfield{author}{\bibinfo{person}{Leigh McAlister} {and}
  \bibinfo{person}{Edgar Pessemier}.} \bibinfo{year}{1982}\natexlab{}.
\newblock \showarticletitle{Variety seeking behavior: An interdisciplinary
  review}.
\newblock \bibinfo{journal}{\emph{Journal of Consumer research}}
  \bibinfo{volume}{9}, \bibinfo{number}{3} (\bibinfo{year}{1982}),
  \bibinfo{pages}{311--322}.
\newblock


\bibitem[\protect\citeauthoryear{McNee, Riedl, and Konstan}{McNee
  et~al\mbox{.}}{2006}]%
        {mcnee2006being}
\bibfield{author}{\bibinfo{person}{Sean~M McNee}, \bibinfo{person}{John Riedl},
  {and} \bibinfo{person}{Joseph~A Konstan}.} \bibinfo{year}{2006}\natexlab{}.
\newblock \showarticletitle{Being accurate is not enough: how accuracy metrics
  have hurt recommender systems}. In \bibinfo{booktitle}{\emph{CHI'06 extended
  abstracts on Human factors in computing systems}}.
  \bibinfo{pages}{1097--1101}.
\newblock


\bibitem[\protect\citeauthoryear{Murakami, Mori, and Orihara}{Murakami
  et~al\mbox{.}}{2007}]%
        {murakami2007metrics}
\bibfield{author}{\bibinfo{person}{Tomoko Murakami}, \bibinfo{person}{Koichiro
  Mori}, {and} \bibinfo{person}{Ryohei Orihara}.}
  \bibinfo{year}{2007}\natexlab{}.
\newblock \showarticletitle{Metrics for evaluating the serendipity of
  recommendation lists}. In \bibinfo{booktitle}{\emph{Annual conference of the
  Japanese society for artificial intelligence}}. Springer,
  \bibinfo{pages}{40--46}.
\newblock


\bibitem[\protect\citeauthoryear{Nguyen, Hui, Harper, Terveen, and
  Konstan}{Nguyen et~al\mbox{.}}{2014}]%
        {nguyen2014exploring}
\bibfield{author}{\bibinfo{person}{Tien~T Nguyen}, \bibinfo{person}{Pik-Mai
  Hui}, \bibinfo{person}{F~Maxwell Harper}, \bibinfo{person}{Loren Terveen},
  {and} \bibinfo{person}{Joseph~A Konstan}.} \bibinfo{year}{2014}\natexlab{}.
\newblock \showarticletitle{Exploring the filter bubble: the effect of using
  recommender systems on content diversity}. In
  \bibinfo{booktitle}{\emph{Proceedings of the 23rd international conference on
  World wide web}}. ACM, \bibinfo{pages}{677--686}.
\newblock


\bibitem[\protect\citeauthoryear{Pariser}{Pariser}{2011}]%
        {pariser2011filter}
\bibfield{author}{\bibinfo{person}{Eli Pariser}.}
  \bibinfo{year}{2011}\natexlab{}.
\newblock \bibinfo{booktitle}{\emph{The filter bubble: How the new personalized
  web is changing what we read and how we think}}.
\newblock \bibinfo{publisher}{Penguin}.
\newblock


\bibitem[\protect\citeauthoryear{Park and Tuzhilin}{Park and Tuzhilin}{2008}]%
        {park2008long}
\bibfield{author}{\bibinfo{person}{Yoon-Joo Park} {and}
  \bibinfo{person}{Alexander Tuzhilin}.} \bibinfo{year}{2008}\natexlab{}.
\newblock \showarticletitle{The long tail of recommender systems and how to
  leverage it}. In \bibinfo{booktitle}{\emph{Proceedings of the 2008 ACM
  conference on Recommender systems}}. ACM, \bibinfo{pages}{11--18}.
\newblock


\bibitem[\protect\citeauthoryear{Qu, Cai, Ren, Zhang, Yu, Wen, and Wang}{Qu
  et~al\mbox{.}}{2016}]%
        {qu2016product}
\bibfield{author}{\bibinfo{person}{Yanru Qu}, \bibinfo{person}{Han Cai},
  \bibinfo{person}{Kan Ren}, \bibinfo{person}{Weinan Zhang},
  \bibinfo{person}{Yong Yu}, \bibinfo{person}{Ying Wen}, {and}
  \bibinfo{person}{Jun Wang}.} \bibinfo{year}{2016}\natexlab{}.
\newblock \showarticletitle{Product-based neural networks for user response
  prediction}. In \bibinfo{booktitle}{\emph{2016 IEEE 16th International
  Conference on Data Mining (ICDM)}}. IEEE, \bibinfo{pages}{1149--1154}.
\newblock


\bibitem[\protect\citeauthoryear{Rumelhart, Hinton, and Williams}{Rumelhart
  et~al\mbox{.}}{1985}]%
        {rumelhart1985learning}
\bibfield{author}{\bibinfo{person}{David~E Rumelhart},
  \bibinfo{person}{Geoffrey~E Hinton}, {and} \bibinfo{person}{Ronald~J
  Williams}.} \bibinfo{year}{1985}\natexlab{}.
\newblock \bibinfo{booktitle}{\emph{Learning internal representations by error
  propagation}}.
\newblock \bibinfo{type}{{T}echnical {R}eport}.
  \bibinfo{institution}{California Univ San Diego La Jolla Inst for Cognitive
  Science}.
\newblock


\bibitem[\protect\citeauthoryear{Sedhain, Menon, Sanner, and Xie}{Sedhain
  et~al\mbox{.}}{2015}]%
        {sedhain2015autorec}
\bibfield{author}{\bibinfo{person}{Suvash Sedhain},
  \bibinfo{person}{Aditya~Krishna Menon}, \bibinfo{person}{Scott Sanner}, {and}
  \bibinfo{person}{Lexing Xie}.} \bibinfo{year}{2015}\natexlab{}.
\newblock \showarticletitle{Autorec: Autoencoders meet collaborative
  filtering}. In \bibinfo{booktitle}{\emph{Proceedings of the 24th
  International Conference on World Wide Web}}. ACM, \bibinfo{pages}{111--112}.
\newblock


\bibitem[\protect\citeauthoryear{Shani and Gunawardana}{Shani and
  Gunawardana}{2011}]%
        {shani2011evaluating}
\bibfield{author}{\bibinfo{person}{Guy Shani} {and} \bibinfo{person}{Asela
  Gunawardana}.} \bibinfo{year}{2011}\natexlab{}.
\newblock \showarticletitle{Evaluating recommendation systems}.
\newblock In \bibinfo{booktitle}{\emph{Recommender systems handbook}}.
  \bibinfo{publisher}{Springer}, \bibinfo{pages}{257--297}.
\newblock


\bibitem[\protect\citeauthoryear{Shaw, Uszkoreit, and Vaswani}{Shaw
  et~al\mbox{.}}{2018}]%
        {shaw2018self}
\bibfield{author}{\bibinfo{person}{Peter Shaw}, \bibinfo{person}{Jakob
  Uszkoreit}, {and} \bibinfo{person}{Ashish Vaswani}.}
  \bibinfo{year}{2018}\natexlab{}.
\newblock \showarticletitle{Self-attention with relative position
  representations}.
\newblock \bibinfo{journal}{\emph{arXiv preprint arXiv:1803.02155}}
  (\bibinfo{year}{2018}).
\newblock


\bibitem[\protect\citeauthoryear{Shi, Hu, Zhao, and Yu}{Shi
  et~al\mbox{.}}{2018}]%
        {shi2018heterogeneous}
\bibfield{author}{\bibinfo{person}{Chuan Shi}, \bibinfo{person}{Binbin Hu},
  \bibinfo{person}{Xin Zhao}, {and} \bibinfo{person}{Philip Yu}.}
  \bibinfo{year}{2018}\natexlab{}.
\newblock \showarticletitle{Heterogeneous Information Network Embedding for
  Recommendation}.
\newblock \bibinfo{journal}{\emph{IEEE Transactions on Knowledge and Data
  Engineering}} (\bibinfo{year}{2018}).
\newblock


\bibitem[\protect\citeauthoryear{Shi, Li, Zhang, Sun, and Philip}{Shi
  et~al\mbox{.}}{2017}]%
        {shi2017survey}
\bibfield{author}{\bibinfo{person}{Chuan Shi}, \bibinfo{person}{Yitong Li},
  \bibinfo{person}{Jiawei Zhang}, \bibinfo{person}{Yizhou Sun}, {and}
  \bibinfo{person}{S~Yu Philip}.} \bibinfo{year}{2017}\natexlab{}.
\newblock \showarticletitle{A survey of heterogeneous information network
  analysis}.
\newblock \bibinfo{journal}{\emph{IEEE Transactions on Knowledge and Data
  Engineering}} \bibinfo{volume}{29}, \bibinfo{number}{1}
  (\bibinfo{year}{2017}), \bibinfo{pages}{17--37}.
\newblock


\bibitem[\protect\citeauthoryear{Snoek, Larochelle, and Adams}{Snoek
  et~al\mbox{.}}{2012}]%
        {snoek2012practical}
\bibfield{author}{\bibinfo{person}{Jasper Snoek}, \bibinfo{person}{Hugo
  Larochelle}, {and} \bibinfo{person}{Ryan~P Adams}.}
  \bibinfo{year}{2012}\natexlab{}.
\newblock \showarticletitle{Practical bayesian optimization of machine learning
  algorithms}. In \bibinfo{booktitle}{\emph{Advances in neural information
  processing systems}}. \bibinfo{pages}{2951--2959}.
\newblock


\bibitem[\protect\citeauthoryear{Wang, Cao, and Wang}{Wang
  et~al\mbox{.}}{2019}]%
        {wang2019survey}
\bibfield{author}{\bibinfo{person}{Shoujin Wang}, \bibinfo{person}{Longbing
  Cao}, {and} \bibinfo{person}{Yan Wang}.} \bibinfo{year}{2019}\natexlab{}.
\newblock \showarticletitle{A survey on session-based recommender systems}.
\newblock \bibinfo{journal}{\emph{arXiv preprint arXiv:1902.04864}}
  (\bibinfo{year}{2019}).
\newblock


\bibitem[\protect\citeauthoryear{Zhang, Yao, Sun, and Tay}{Zhang
  et~al\mbox{.}}{2019}]%
        {zhang2019deep}
\bibfield{author}{\bibinfo{person}{Shuai Zhang}, \bibinfo{person}{Lina Yao},
  \bibinfo{person}{Aixin Sun}, {and} \bibinfo{person}{Yi Tay}.}
  \bibinfo{year}{2019}\natexlab{}.
\newblock \showarticletitle{Deep learning based recommender system: A survey
  and new perspectives}.
\newblock \bibinfo{journal}{\emph{ACM Computing Surveys (CSUR)}}
  \bibinfo{volume}{52}, \bibinfo{number}{1} (\bibinfo{year}{2019}),
  \bibinfo{pages}{5}.
\newblock


\bibitem[\protect\citeauthoryear{Zhang, S{\'e}aghdha, Quercia, and
  Jambor}{Zhang et~al\mbox{.}}{2012}]%
        {zhang2012auralist}
\bibfield{author}{\bibinfo{person}{Yuan~Cao Zhang},
  \bibinfo{person}{Diarmuid~{\'O} S{\'e}aghdha}, \bibinfo{person}{Daniele
  Quercia}, {and} \bibinfo{person}{Tamas Jambor}.}
  \bibinfo{year}{2012}\natexlab{}.
\newblock \showarticletitle{Auralist: introducing serendipity into music
  recommendation}. In \bibinfo{booktitle}{\emph{Proceedings of the fifth ACM
  international conference on Web search and data mining}}. ACM,
  \bibinfo{pages}{13--22}.
\newblock


\bibitem[\protect\citeauthoryear{Zheng, Chan, and Ip}{Zheng
  et~al\mbox{.}}{2015}]%
        {zheng2015unexpectedness}
\bibfield{author}{\bibinfo{person}{Qianru Zheng}, \bibinfo{person}{Chi-Kong
  Chan}, {and} \bibinfo{person}{Horace~HS Ip}.}
  \bibinfo{year}{2015}\natexlab{}.
\newblock \showarticletitle{An unexpectedness-augmented utility model for
  making serendipitous recommendation}. In \bibinfo{booktitle}{\emph{Industrial
  Conference on Data Mining}}. Springer, \bibinfo{pages}{216--230}.
\newblock


\bibitem[\protect\citeauthoryear{Zhou, Mou, Fan, Pi, Bian, Zhou, Zhu, and
  Gai}{Zhou et~al\mbox{.}}{2019}]%
        {zhou2019deep}
\bibfield{author}{\bibinfo{person}{Guorui Zhou}, \bibinfo{person}{Na Mou},
  \bibinfo{person}{Ying Fan}, \bibinfo{person}{Qi Pi}, \bibinfo{person}{Weijie
  Bian}, \bibinfo{person}{Chang Zhou}, \bibinfo{person}{Xiaoqiang Zhu}, {and}
  \bibinfo{person}{Kun Gai}.} \bibinfo{year}{2019}\natexlab{}.
\newblock \showarticletitle{Deep interest evolution network for click-through
  rate prediction}. In \bibinfo{booktitle}{\emph{Proceedings of the AAAI
  Conference on Artificial Intelligence}}, Vol.~\bibinfo{volume}{33}.
  \bibinfo{pages}{5941--5948}.
\newblock


\bibitem[\protect\citeauthoryear{Zhou, Zhu, Song, Fan, Zhu, Ma, Yan, Jin, Li,
  and Gai}{Zhou et~al\mbox{.}}{2018}]%
        {zhou2018deep}
\bibfield{author}{\bibinfo{person}{Guorui Zhou}, \bibinfo{person}{Xiaoqiang
  Zhu}, \bibinfo{person}{Chenru Song}, \bibinfo{person}{Ying Fan},
  \bibinfo{person}{Han Zhu}, \bibinfo{person}{Xiao Ma},
  \bibinfo{person}{Yanghui Yan}, \bibinfo{person}{Junqi Jin},
  \bibinfo{person}{Han Li}, {and} \bibinfo{person}{Kun Gai}.}
  \bibinfo{year}{2018}\natexlab{}.
\newblock \showarticletitle{Deep interest network for click-through rate
  prediction}. In \bibinfo{booktitle}{\emph{Proceedings of the 24th ACM SIGKDD
  International Conference on Knowledge Discovery \& Data Mining}}. ACM,
  \bibinfo{pages}{1059--1068}.
\newblock


\bibitem[\protect\citeauthoryear{Zhou, Kuscsik, Liu, Medo, Wakeling, and
  Zhang}{Zhou et~al\mbox{.}}{2010}]%
        {zhou2010solving}
\bibfield{author}{\bibinfo{person}{Tao Zhou}, \bibinfo{person}{Zolt{\'a}n
  Kuscsik}, \bibinfo{person}{Jian-Guo Liu}, \bibinfo{person}{Mat{\'u}{\v{s}}
  Medo}, \bibinfo{person}{Joseph~Rushton Wakeling}, {and}
  \bibinfo{person}{Yi-Cheng Zhang}.} \bibinfo{year}{2010}\natexlab{}.
\newblock \showarticletitle{Solving the apparent diversity-accuracy dilemma of
  recommender systems}.
\newblock \bibinfo{journal}{\emph{Proceedings of the National Academy of
  Sciences}} \bibinfo{volume}{107}, \bibinfo{number}{10}
  (\bibinfo{year}{2010}), \bibinfo{pages}{4511--4515}.
\newblock


\bibitem[\protect\citeauthoryear{Ziegler, McNee, Konstan, and Lausen}{Ziegler
  et~al\mbox{.}}{2005}]%
        {ziegler2005improving}
\bibfield{author}{\bibinfo{person}{Cai-Nicolas Ziegler},
  \bibinfo{person}{Sean~M McNee}, \bibinfo{person}{Joseph~A Konstan}, {and}
  \bibinfo{person}{Georg Lausen}.} \bibinfo{year}{2005}\natexlab{}.
\newblock \showarticletitle{Improving recommendation lists through topic
  diversification}. In \bibinfo{booktitle}{\emph{Proceedings of the 14th
  international conference on World Wide Web}}. ACM, \bibinfo{pages}{22--32}.
\newblock


\end{thebibliography}
\end{document}